\documentclass[10pt,a4paper,final]{iopart}
\expandafter\let\csname equation*\endcsname\relax
\expandafter\let\csname endequation*\endcsname\relax
\usepackage[margin=1 in]{geometry}
\usepackage{amsmath,color,iopams,graphicx}
\usepackage[breaklinks=true,colorlinks=true,linkcolor=blue,urlcolor=blue,citecolor=blue]{hyperref}
\usepackage[english]{babel}
\usepackage{graphicx,comment}
\usepackage{mathtools}
\usepackage{amsmath}
\usepackage{footnote}
\makesavenoteenv{tabular}
\makesavenoteenv{table}
\usepackage{multicols}
\usepackage{subfigure}
\usepackage{widetext}
\usepackage{caption}
\usepackage{float}
\begin{document}
\title{Density-density correlation functions of chiral Luttinger liquids with a point-contact impurity}
\author{  Nikhil Danny Babu$^{\dagger}$\footnote{Corresponding Author}, Girish S. Setlur$^{*}$}
\address{Department of Physics \\ Indian Institute of Technology  Guwahati \\ Guwahati, Assam 781039, India}
\ead{$^{\dagger}$danny@iitg.ac.in,$^{*}$gsetlur@iitg.ernet.in}

\begin{abstract}
The density-density correlation functions (most singular parts) of chiral Luttinger liquids forming the fractional quantum Hall effect (FQHE) edge are systematically derived in presence of a point-contact junction acting as a localised scalar impurity and are shown to be expressible as compact analytical functions with second order poles and involving the bare scale-independent reflection and transmission coefficients. The results are validated on comparison with standard fermionic perturbation theory. The linear response Hall conductance in the absence of a point-contact is recovered from the obtained density-density correlation functions (DDCF). The system under consideration is inhomogeneous with broken translational invariance and for such systems in one dimension, the connected moments of the density fluctuation operator beyond second order need not be included in order to retrieve the most singular parts of the correlations. The reason being that all odd moments of the density are zero and all higher order even moments are less singular than the quadratic moment. The implications of these results when used in conjunction with bosonization methods in presence of impurity backscattering is briefly discussed.
\end{abstract}

\vspace{2pc}
\section{Introduction}
The study of interacting one-dimensional (1D) systems has been a prominent and active area of research in condensed matter physics. It is known that interacting systems in one dimension exhibit a Luttinger liquid phase. Several important analytical \cite{moon1993resonant,kane1992resonant,kane1992transmission,kane1992transport,PhysRevLett.71.3351} and numerical \cite{qin1997impurity,hamamoto2008numerical,freyn2011numerical,Ejima_2009,PhysRevB.99.121103,PhysRevLett.116.247204} studies have given us much insight into the physics of strongly correlated systems in 1D. Edge states that form at the boundaries of a 2D electron gas subjected to strong magnetic fields in the quantum Hall regime are an important example of one-dimensional electron systems. The important realization that the edge excitations of a fractional quantum Hall effect (FQHE) system could be described as 1D chiral Luttinger liquids was due to Wen \cite{PhysRevB.43.11025}. An FQHE bulk with filling factor $\nu = 1/m$ where $m$ is an odd integer (the Laughlin series) has a single edge mode that can be treated as a chiral Luttinger liquid. Fractional quantum Hall edge states are the most experimentally feasible systems to probe Luttinger liquid physics. Although the edge state fermions are topologically protected by the bulk and do not backscatter, a point-contact constriction in the bulk brings the opposite edges close together and the chiral fermions can backscatter across this impurity. Thus the quantum Hall system is well suited to study backscattering in Luttinger liquids since it is easy to create isolated impurities and probe the transport properties experimentally. The ideal way to deal with interactions in one dimension is through bosonization \cite{haldane1981luttinger}. The density-density correlation functions are necessary to obtain the Green's functions of the Luttinger liquid using bosonization. Conventional bosonization methods for chiral Luttinger liquids \cite{doi:https://doi.org/10.1002/9783527617258.ch4} is not well suited to study backscattering from an isolated impurity. Often the procedure is to deal with the impurity in a perturbative manner. Novel bosonization techniques that modify the Fermi-Bose correspondence to include the impurity backscattering exactly have proven to be successful in studying Luttinger liquids with impurities \cite{doi:10.1142/S0217751X18501749,Das_2018,Danny_Babu_2020,Das_2020}. Application of such bosonization techniques to study chiral Luttinger liquids coupled through a point-contact would require the density-density correlation functions (DDCF) in presence of impurity backscattering.\\
 Several important works involving the density-density correlation in 1D can be found in the literature. The density correlations in presence of long range Coulomb forces in a one dimensional electron gas was investigated by Schulz \cite{schulz1993wigner}. The zero temperature Fourier transform of the $2 k _{F}$ component of the DDCF of a Luttinger liquid with spinful fermions was obtained in \cite{iucci2007fourier}. The density-density correlation function for the half-filled Hubbard model with nearest-neighbour repulsion and large on-site repulsion compared to hopping amplitudes was calculated in \cite{stephan1996dynamical}. Density correlations of Luttinger liquids with curvature in the dispersion was explored by Aristov in \cite{aristov2007luttinger}. The four-point correlations of a Luttinger liquid in a nonequilibrium setting was investigated by Protopopov et al.\cite{protopopov2011many}. Numerical studies of the DDCF in the lattice Calogero Sutherland model showing Luttinger liquid behaviour was performed in \cite{sen1999density}.\\\mbox{ }\\
 In this work we evaluate the most singular parts of the density-density correlation functions (DDCF) of chiral Luttinger liquids (assumed spinless) with a point-contact junction at the origin. This is isomorphic to a system of fractional quantum Hall effect (FQHE) edge states belonging to the Laughlin series (with filling fraction $\nu = 1/m$ with $m$ odd) with a point-contact constriction. The edge modes are chiral in the sense that electron transport is unidirectional along the edges. If we consider a square geometry of the FQHE bulk, the opposing edge states will have opposite chirality and this chirality is protected by the topology of the bulk. The basic feature of the Luttinger liquid model is the introduction of two species of fermions (left movers and right movers). The right movers have an energy spectrum $\epsilon_{k} = \hbar v_{F} k$ and the left movers have an energy spectrum given by $\epsilon_{k} = -\hbar v_{F} k$, where $v_{F}$ is the Fermi velocity, $k$ is the wavevector and $\hbar$ is the reduced Planck's constant ($\hbar = 1$ is assumed in the rest of the paper). For an ordinary Luttinger liquid the right and left movers are present in the same one-dimensional chain whereas in a chiral Luttinger liquid only the right ($R$) movers or left ($L$) movers are present. Chiral Luttinger liquids are always associated with a $2+1 \mbox{ }D$ bulk and they cannot exist independent of the bulk. In Laughlin fractional quantum Hall systems the right ($R$) and left ($L$) moving edges are spatially separated by the bulk and form chiral Luttinger liquids and the right and left movers are coupled through an interaction mediated by the bulk. In the following sections the most singular part of the DDCF is obtained using a generating functional method and the results are compared with standard perturbation theory. The paper is organized as follows: In Sec.\ref{secmodel} we introduce the model Hamiltonian and discuss the density-density correlations of the non-interacting model. In Sec.\ref{genfunc} the generating functional approach to calculate the density correlations in presence of interactions is discussed. The detailed calculations of the correlation functions are shown in Secs.\ref{symmetriccorrel} and \ref{antisymmetriccorrel} and the final results in a compact form are shown in Sec.\ref{finresults}. A term by term comparison of our results with standard perturbation theory is done in Sec.\ref{pertcomparison}. In Sec.\ref{linearcurrent} we calculate the linear response current and conductance due to a potential difference between the right and left mover edges and obtain the expected result of the two-terminal Hall conductance in absence of a point-contact and obtain a backscattering term in presence of a point-contact at zero temperature. Discussion and summarization of the main results of this work is done in Sec.\ref{conclusions}.

\section{Model Hamiltonian}
\label{secmodel}
We consider a system of two chiral Luttinger liquids with a point-contact tunnel junction at the origin. This can be realized as edge states of a fractional quantum Hall effect (FQHE) fluid with a narrow constriction in the middle that forms a point-contact between the opposite edges. The FQHE systems belonging to the Laughlin series \cite{PhysRevLett.50.1395} with filling fraction $\nu=1/m$ where $m$ is an odd integer have only single edge modes that can be treated as chiral Luttinger liquids \cite{PhysRevB.41.12838}. The non-interacting part of the Hamiltonian is
\begin{align}
H_{0} = \sum_{p}(v_F p)c^{\dagger}_{p,R}c_{p,R} &+ \sum_{p}(-v_Fp)c^{\dagger}_{p,L}c_{p,L}+ \frac{\Gamma}{L}(c^{\dagger}_{.,R}c_{.,L}+c^{\dagger}_{.,L}c_{.,R})
 \label{eqhamnonint}
\end{align}
where $R$ and $L$ label the right and left moving spinless fermions respectively. The first two terms of the Hamiltonian constitute the kinetic part and the last term denotes the point-contact tunneling. We are only interested in the low-energy physics, hence we consider a linearized dispersion close to the right and left Fermi points. The fermion creation and annihilation operators in momentum space are $c^{\dagger}_{p}$ and $c_{p}$ respectively and we use the notation $c^{\dagger}_{.,R} = \sum_{p}c^{\dagger}_{p,R}$ in the point-contact tunneling term of the Hamiltonian. The point-contact is considered to be at the origin in real space. This is the equilibrium version of the Hamiltonian studied in \cite{Babu_2022}. The tunneling amplitude of the symmetric point-contact junction is $\Gamma$ and $L$ is the system size. The tunneling term is the spinless and symmetric version of the backscattering part of Eq.3 in \cite{Danny_Babu_2020}. The interaction part of the Hamiltonian is
\begin{align}
H_{int} = \int dx\mbox{ } v_{0}\mbox{ } \rho_{R}(x,t) \rho_{L}(x,t)
\label{hint}
\end{align}
The right mover and left mover densities are coupled through a bulk mediated interaction parametrized by $v_{0}$. For chiral Luttinger liquids of opposite chirality coupled through a fractional quantum Hall bulk, the interaction parameter is expressed in terms of the bulk filling fraction $\nu$. This interaction term is identical to the $g_{2}$ term in the standard g-ology model \cite{giamarchi2003quantum} which describes forward scattering density-density interactions between the different species of fermions and is what gives rise to Luttinger liquid physics. The interaction term that couples the same species of fermions to each other merely shifts the fermion velocity and does not directly contribute to Luttinger liquid behaviour and hence it is not included in our Hamiltonian. The full Hamiltonian of interest is
\begin{align}
H\mbox{ }=\mbox{ }H_{0} + H_{int}
\end{align}
The exact Green's functions in space-time domain for the non-interacting problem in presence of a voltage bias between the right and left movers was obtained using analytic methods in \cite{Babu_2022}. In this work we are interested in the equilibrium situation without the bias. The finite temperature Green's function in equilibrium is periodic in imaginary time. The time variable when rotated onto the imaginary axis is periodic in the interval $[0, -i \beta]$ where $\beta = \frac{1}{T}$ is the inverse of the temperature $T$ of the system. The equilibrium non-interacting density-density correlation functions (DDCF) are obtained from the two-point Green's functions using Wick's theorem and are written as \cite{Babu_2022},
 \begin{align}
  <&T \mbox{     } \rho_R(x,t) \rho_R(x',t') >_0 -<\rho_R(x,t) >_0<\rho_R(x',t') >_0\mbox{          }
 \nonumber \\ &=  \mbox{          }<T \mbox{     } \rho_L(-x,t) \rho_L(-x',t') >_0-<\rho_L(-x,t) >_0<\rho_L(-x',t') >_0 \mbox{          }\nonumber \\
&=\mbox{ } \left[    \frac{i}{2\pi} \frac{ \frac{ \pi }{\beta v_F } }{\sinh( \frac{ \pi }{\beta v_F } (x-x'-v_F(t-t') ) )  } \right]^2 \nonumber \\ &
\mbox{          }\mbox{          }\mbox{          }\mbox{          }\mbox{          }
    \bigg(   \left[1
-      \theta(x')    \mbox{          }   \frac{  2 \Gamma^2
}{\Gamma ^2 +4 v_F^2}\right]^2\mbox{          }    \left[1
-      \theta(x )    \mbox{          }   \frac{  2 \Gamma^2
}{\Gamma ^2 +4 v_F^2}\right]^2
  + \left( \frac{\Gamma }{v_F} \mbox{          } \frac{(2 v_F)^2
}{\Gamma ^2 +4 v_F^2}\right)^4 \mbox{          }
  \theta(x )     \theta(x')  \nonumber \\
  &\mbox{ }\mbox{          }\mbox{          }\mbox{          }+
\mbox{          }
2\left[1
-       \frac{  2 \Gamma^2
}{\Gamma ^2 +4 v_F^2}\right]^2 \mbox{          }   \left( \frac{\Gamma }{v_F} \mbox{          } \frac{(2 v_F)^2
}{\Gamma ^2 +4 v_F^2}\right)^2 \mbox{          } \theta(x )     \theta(x' )   \mbox{          }
   \bigg)
  \label{rhorhoRRLL}
 \end{align}
 and
 \begin{align}
<&T \mbox{     } \rho_R(x,t) \rho_L(-x',t') >_0-<\rho_R(x,t)>_0< \rho_L(-x',t') >_0 \mbox{          }
 \nonumber \\ &=  \mbox{          }<T \mbox{     } \rho_L(-x,t) \rho_R(x',t') >_0 - <\rho_L(-x,t) >_0<\rho_R(x',t') >_0\mbox{          }\nonumber \\
 &= \mbox{ }\left[  \frac{i}{2\pi} \frac{ \frac{ \pi }{\beta v_F } }{\sinh( \frac{ \pi }{\beta v_F } (x-x'-v_F(t-t') ) )  }\right]^2 \nonumber \\ &
 \mbox{          }\mbox{          }\mbox{          }\mbox{          }\mbox{          }
 \bigg(\left( i  \frac{\Gamma }{v_F}\frac{(2 v_F)^2
}{\Gamma ^2 +4 v_F^2} \right)^2
\mbox{          }
  \left( -    \left[ 1
-      \theta(x )    \mbox{          }   \frac{  2 \Gamma^2
}{\Gamma ^2 +4 v_F^2}
\right]^2    \theta(x')  -
 \left[1
-      \theta(x' )    \mbox{          }   \frac{  2 \Gamma^2
}{\Gamma ^2 +4 v_F^2}\right]^2    \theta( x)      \right) \nonumber \\
&\mbox{          }\mbox{          }\mbox{          }\mbox{          }+  \left( i  \frac{\Gamma }{v_F}\frac{(2 v_F)^2
}{\Gamma ^2 +4 v_F^2} \right)^2
\mbox{          }
  2 \mbox{          }     \left[ 1
-       \frac{  2 \Gamma^2
}{\Gamma ^2 +4 v_F^2}
\right]^2  \theta(x)
  \theta( x' )\bigg)
  \label{rhorhoRLLR}
 \end{align}
 where $v_{F}$ is the Fermi velocity, $\beta$ is the inverse temperature and $\theta(x)$ is the Dirichlet regularized step function, the precise definition of which is explained in the next paragraph. 
The density-density correlations satisfy the following identities,
\begin{align}
 <&T \mbox{     } \rho_{\chi}(\chi \mbox{ }x,t) (\rho_{\chi}(\chi\mbox{ }x',t')+\rho_{-\chi}(-\chi \mbox{ }x',t')) >_0 -<\rho_{\chi}(x,t) >_0<(\rho_{\chi}(\chi \mbox{ }x',t')+\rho_{-\chi}(-\chi\mbox{ }x',t')) >_0\mbox{          }=
 \mbox{          }\nonumber \\
  <&T \mbox{     }(\rho_{\chi}(\chi \mbox{ }x,t) + \rho_{-\chi}(-\chi \mbox{ }x,t)  ) \rho_{\chi}(\chi \mbox{ }x',t') >_0 - <(\rho_{\chi}(\chi \mbox{ }x,t) + \rho_{-\chi}(-\chi \mbox{ }x,t) ) >_0<\rho_{\chi}(\chi\mbox{ }x',t') >_0 \nonumber \\ &\mbox{     }=  \mbox{          }
  \left[    \frac{i}{2\pi} \frac{ \frac{ \pi }{\beta v_F } }{\sinh( \frac{ \pi }{\beta v_F } (x-x'-v_F(t-t') ) )  } \right]^2
  \label{denidentity}
 \end{align}
 where $\chi$ takes values $\pm 1$ with $1$ denoting $R$ (right movers) and $-1$ denoting $L$ (left movers). The regularization scheme implied in writing Eq.\ref{eqhamnonint} is that summimg over all momenta implicitly means summing over momenta in the interval $p \in \lbrace-\Lambda,\Lambda\rbrace $ and taking $\Lambda \rightarrow \infty$. The implication of this Dirichlet regularization scheme is that the values of discontinuous functions are always the average of the left and right hand limits. The step function appearing throughout this paper is the Dirichlet regularized step function defined as: $\theta(x>0) = 1,\mbox{ }\theta(x<0)=0,\mbox{ }\theta(x=0)=\frac{1}{2}$. In presence of the interactions it is in general not possible to write down a simple formula such as in Eqs.\ref{rhorhoRRLL} and \ref{rhorhoRLLR} for the correlation functions when the point-contact is present. There is no guarantee that the correlation function will have simple second order poles. However it is possible to obtain the most singular part of the density-density correlation function even in the presence of forward scattering interactions, that is when Eq.\ref{hint} is present in the Hamiltonian. By most singular we mean terms with higher order poles (second order in this case). In the following sections we derive the most singular contribution to the interacting DDCF using a generating functional. Before proceeding let us make the following definition of the symmetric and antisymmetric density (fluctuation) operators
 \begin{align}
 &\rho_{sym}(x,t) \equiv \rho_{R}(x,t)+\rho_{L}(-x,t) \mbox{ };\nonumber \\
 &\rho_{asy}(x,t) \equiv \rho_{R}(x,t)-\rho_{L}(-x,t)
 \label{rsymnonint}
 \end{align}
 This means we can write,
 \begin{align}
 \rho_{R}(x,t) = \frac{\rho_{sym}(x,t)+\rho_{asy}(x,t) }{2}
 \end{align}
 and
 \begin{align}
 \rho_{L}(-x,t) = \frac{\rho_{sym}(x,t)-\rho_{asy}(x,t) }{2}
 \end{align}
 Using Eqs.\ref{rhorhoRRLL},\ref{rhorhoRLLR} and \ref{denidentity} we express the non-interacting correlations of the symmetric and antisymmetric densities as,
 \begin{align}
 <\rho_{sym}(x,t)\rho_{sym}(x',t')>_{0} \mbox{ }=\mbox{ }2\left[    \frac{i}{2\pi} \frac{ \frac{ \pi }{\beta v_F } }{\sinh( \frac{ \pi }{\beta v_F } (x-x'-v_F(t-t') ) )  } \right]^2
 \label{rsymrsym}
 \end{align}
\begin{align}
 <\rho_{sym}(x,t)\rho_{asy}(x',t')>_{0} \mbox{ }=\mbox{ } 0
 \label{rhosymasy0}
 \end{align}
 \begin{align}
 <\rho_{asy}(x,t)\rho_{sym}(x',t')>_{0} \mbox{ }=\mbox{ } 0
 \label{rhoasysym0}
 \end{align}
  \begin{align}
 <\rho_{asy}(x,t)\rho_{asy}(x',t')>_{0} \mbox{ }=\mbox{ }(\theta(xx') + r_1 \mbox{  }\theta(-xx')) \mbox{   }<T\rho_{sym}(x,t)\rho_{sym}(x',t')>_{0}
 \end{align}
where $r_1 = \left(1
  -\frac{32 \Gamma ^2 v_F^2}{\left(\Gamma ^2+4 v_F^2\right)^2}
\right)$. The densities by definition are precisely the density fluctuations ($\rho \mbox{ }-\mbox{ } \rho_{0}$ where $\rho_{0}$ is the average density) and all the correlation functions discussed in this work are by default the connected correlations. We choose to work with the $\rho_{sym}$ and $\rho_{asy}$ fields as they prove to be convenient in evaluating the path integrals in the generating functional. The interaction part of the Hamiltonian is expressed in terms of these new fields as
\begin{align}
H_{int} = \int dx \mbox{       } v_0 \mbox{     } \rho_R(x,t) \rho_L(x,t)  =\frac{ v_0 }{4}\mbox{     }  \int dx \mbox{       }  (\rho_{sym}(x,t)+\rho_{asy}(x,t) )
 ( \rho_{sym}(-x,t)-\rho_{asy}(-x,t) )
 \label{hintsymasy}
\end{align}
\section{Generating functional for the density-density correlations}
\label{genfunc}
Interparticle interactions are systematically introduced non-perturbatively in the density correlations by means of a generating functional approach. In our model the curvature in the Fermi surface is neglected and we only focus on the low-energy physics, which means that we deal with excitations close to the Fermi surface. The dispersion is linearized close to the Fermi point. In other words we are working in the random phase approximation (RPA) limit \cite{stone1994bosonization,dzyaloshinskii1974correlation} where the Fermi momentum and mass of the fermion is allowed to diverge in such a way that their ratio is finite (i.e. $k_{F},\mbox{ }m \rightarrow \infty$ but $k_{F}/m = v_{F} < \infty$). It is possible to obtain a compact analytical expression for the density correlations in presence of an arbitrary impurity provided we restrict to the most singular contributions only. This idea relies on a careful redefinition of the RPA for inhomogeneous systems (this is discussed in \hyperref[AppendixD]{Appendix D}). This redefinition involves a systematic truncation and resummation of the perturbation series in powers of the fermion-fermion coupling with the impurities being arbitrary. Only when such a truncation and resummation is employed, closed analytical expressions for the correlation functions is possible for a system as complicated as mutually interacting fermions in presence of impurities. The generating functional we employ to calculate the $<\rho_{sym}\rho_{sym}>$ and $<\rho_{asy}\rho_{asy}>$ correlations with auxiliary fields $U_{sym}$ and $U_{asy}$ in presence of interactions and the point-contact impurity can be written as
\begin{align}
 Z[U] = \int D[\rho_{sym} ] \int D[\rho_{asy}]\mbox{ } \mbox{    }  e^{ i S_0 }e^{ i S_{int} } e^{ \int \rho_{sym} U_{sym} +  \int \rho_{asy} U_{asy} }
 \label{zu}
 \end{align}
where $S_{0}$ and $S_{int}$ are the free and interacting actions respectively. The generating functional in the absence of interactions is
\begin{align}
 Z_0[U] = &\int D[\rho_{sym} ] \int D[\rho_{asy}]\mbox{ } \mbox{    }  e^{ i S_0 } e^{ \int \rho_{sym} U_{sym} +  \int \rho_{asy} U_{asy} }\nonumber \\
 &\Rightarrow e^{ i S_0 } = \int D[U^{'}_{sym} ] \int D[U^{'}_{asy}]\mbox{ } \mbox{    }  e^{ -\int \rho_{sym} U^{'}_{sym} -  \int \rho_{asy} U^{'}_{asy} }\mbox{    } \mbox{    } Z_0[U^{'}]
 \label{z0}
 \end{align}
 Hence we can write
 \begin{align}
 Z[U] =  &\int D[U^{'}_{sym} ] \int D[U^{'}_{asy}]\mbox{    } Z_0[U^{'}] \mbox{ } \mbox{    } \int D[\rho_{sym} ] \int D[\rho_{asy}]\nonumber \\&e^{ -i \int_C dt \mbox{          }   \int dx  \mbox{ } \frac{ v_0 }{4}\mbox{     }   (\rho_{sym}(x,t)+\rho_{asy}(x,t) )
 ( \rho_{sym}(-x,t)-\rho_{asy}(-x,t) )   }
\mbox{      }\nonumber \\&e^{  \int_C dt \mbox{          }   \int dx  \mbox{ }  \rho_{sym}(x,t)( U_{sym}(x,t) - U^{'}_{sym}(x,t) )  +  \int_C dt \mbox{          }   \int dx  \mbox{ } \rho_{asy}(x,t) ( U_{asy}(x,t) - U^{'}_{asy}(x,t)) }
\label{zu}
 \end{align}
 since $e^{ i S_{int} }  =  e^{ -i \int_C dt \mbox{          }  H_{int} }$. The path integrals are evaluated using the saddle point method and after integrating over the $\rho_{sym}$ and $\rho_{asy}$ fields we get
 \begin{align}
 Z[U] = &\int D[U^{'}_{sym} ] \int D[U^{'}_{asy}]\mbox{    } Z_0[U^{'}]
 \label{zuonly}
\end{align}
\[
 \times\mbox{ }e^{\substack{ -\frac{1}{i v_{0}} \int_C dt \mbox{  }   \int dx\mbox{ }( U_{asy}(-x,t) - U^{'}_{asy}(-x,t) - U^{'}_{sym}(-x,t) + U_{sym}(-x,t))(U_{asy}(x,t) - U^{'}_{asy}(x,t) + U^{'}_{sym}(x,t) - U_{sym}(x,t))}}
   \]
   We perform the most singular truncation of $Z_{0}$ by making the Gaussian approximation \cite{Danny_Babu_2020}
   \begin{align}
 Z_0[U^{'}] \mbox{ }  = \mbox{    }&e^{ \frac{1}{2} \int_C dt \int dx \int_C dt' \int dx' \mbox{     }<T\rho_{sym}(x,t)\rho_{sym}(x',t')  >_{0,c} \mbox{      }   U^{'}_{sym}(x,t)U^{'}_{sym}(x',t')  }
\mbox{    }\nonumber \\&e^{ \frac{1}{2} \int_C dt \int dx \int_C dt' \int dx' \mbox{     }<T\rho_{asy}(x,t)\rho_{asy}(x',t')  >_{0,c} \mbox{      }   U^{'}_{asy}(x,t)U^{'}_{asy}(x',t')  } 
\label{z0u}
 \end{align}
 In the absence of a point-contact this choice corresponds to RPA. But for an inhomogeneous system this choice corresponds to the most singular truncation of the RPA generating functional. Using the Gaussian approximation implies that the higher connected moments of the density fluctuations beyond quadratic are omitted (a detailed discussion on the validity of the Gaussian approximation is provided in \hyperref[AppendixC]{Appendix C}). We ignore the higher order moments of $\rho$ in $Z_0[U^{'}]$ as they are less singular than the second moment. The connected parts of all the odd moments of $\rho$ vanish and all the higher order even moments are less singular than the second moment as shown in \hyperref[AppendixC]{Appendix C}. The higher order even moments only involve first order simple poles whereas the second moment of the density involves second order poles hence is more singular \cite{Danny_Babu_2020}. This means that this procedure will yield only the most singular part of the DDCF. The most singular part captures essentially the important physics and when used in conjunction with novel bosonization techniques \cite{das2019nonchiral,doi:10.1142/S0217751X18501749,babu2023unconventional} we can obtain the most singular part of the two-point Green's functions of the interacting inhomogeneous system. Solving for the saddle point of the $U^{'}$ fields we obtain the following equations
 \begin{align}
 \int dt' \int dx' \mbox{     }<T\rho_{sym}(x,t)\rho_{sym}(x',t')  >_0 \mbox{      }   U^{'}_{sym}(x',t') + \frac{2i}{ v_0 }\mbox{     }      ( U_{sym}(-x,t) - U^{'}_{sym}(-x,t) )  = 0
 \label{usym}
 \end{align}
 and
 \begin{align}
 \int_C dt' \int dx' \mbox{     }<T\rho_{asy}(x,t)\rho_{asy}(x',t')  >_0 \mbox{      }   U^{'}_{asy}(x',t')      -  \frac{2i}{ v_0 } ( U_{asy}(-x,t) - U^{'}_{asy}(-x,t))= 0
 \label{uasy}
 \end{align}
 The generating functional now in terms of these saddle points is
  \begin{align}
 Z[U] = e^{-\frac{1}{i v_{0}} \int_C dt \int dx \mbox{ } (U^{'}_{sym}(-x,t) - U_{sym}(-x,t)) U_{sym}(x,t) -\frac{1}{i v_{0}} \int_C dt \int dx \mbox{ } (U_{asy}(-x,t) - U^{'}_{asy}(-x,t)) U_{asy}(x,t)}
 \label{zu}
 \end{align}
 We see that the auxiliary fields of the symmetric and antisymmetric densities decouple and hence the $<\rho_{sym}\rho_{sym}>$ and $<\rho_{asy}\rho_{asy}>$ correlations can be separately obtained whereas the $<\rho_{sym}\rho_{asy}>$ correlations are zero.
 \section{Correlation function of the symmetric density fields}
 \label{symmetriccorrel}
 In this section we obtain the correlations of the symmetric densities from the generating functional. In order to proceed we have to obtain the saddle point by solving Eq.\ref{usym} for $U^{'}_{sym}$. It helps to work in the Fourier space. Transforming Eq.\ref{usym} to momentum and Matsubara frequency space we get
 \begin{align}
 -i \beta  \mbox{  }  <\rho_{sym}(q,n)\rho_{sym}(-q,-n)>_0   \mbox{  }
  U^{'}_{sym}(q,n)  + \frac{2i}{ v_0 }   \mbox{  }
 ( U_{sym}(-q,n) -  U^{'}_{sym}(-q,n) ) = 0
 \label{upsymn}
 \end{align}
 where we have used the transformations
\begin{align*}
<T\rho_{sym}(x,t)\rho_{sym}(x',t')>_{0}\mbox{  } = \mbox{  }
      \frac{1}{L}\sum_{q,n} e^{-iq (x-x') } \mbox{ } e^{w_n(t-t') }
      \mbox{        } <\rho_{sym}(q,n)\rho_{sym}(-q,-n)>
\end{align*} 
  and $ U_{sym}(-x,t)  \mbox{        }  =  \mbox{        }\frac{1}{-i \beta L} \sum_{q,n} e^{ i q x } e^{w_n t } \mbox{  }\mbox{  }
  U_{sym}(q,n)$
  and $w_{n} = \frac{2 \pi n}{\beta}$ are bosonic Matsubara frequencies with $n \in \mathbb{Z}$. The saddle point for $U^{'}_{sym}$ is obtained from Eq.\ref{upsymn}.
  \begin{align}
  U^{'}_{sym}(q,n) = \frac{2(v_{0} \beta R_{sym}(-q,n) U_{sym}(-q,n) - 2 U_{sym}(q,n))}{-4+\beta ^2 v_{0}^2 R_{sym}(-q,n) R_{sym}(q,n)}
  \end{align}
  where $R_{sym}(q,n)\mbox{  }=\mbox{   }<\rho_{sym}(q,n)\rho_{sym}(-q,-n)>_0 \mbox{ }=\mbox{ }-\frac{1}{\pi \beta v_{F}}\frac{  i v_F q}{(w_{n}-i v_{F} q)}$ is the non-interacting correlation in momentum-Matsubara frequency space. The generating functional for the $<\rho_{sym} \rho_{sym}>$ correlations is
  \begin{align}
  Z[U_{sym}] &= e^{-  \sum_{q,n}  \mbox{  }\frac{1}{ v_{0} \beta L}
 (U^{'}_{sym}(q,-n) -  U_{sym}(q,-n))
  U_{sym}(q,n) }\nonumber\\
  &= e^{ \frac{1}{  L}\sum_{q,n}
\frac{  R_{sym}(-q,n) (2 U_{sym}(-q,n)-\beta  v_0 R_{sym}(q,n) U_{sym}(q,n))}{4-\beta ^2 v_0^2 R_{sym}(-q,n) R_{sym}(q,n)}
  U_{sym}(q,-n) }
  \end{align}
  Now the correlation functions can be evaluated using the relation
  \begin{align}
&  \frac{1}{  L}\sum_{q,n}
\frac{  R_{sym}(-q,n) (2 U_{sym}(-q,n)-\beta  v_0 R_{sym}(q,n) U_{sym}(q,n))}{4-\beta ^2 v_0^2 R_{sym}(-q,n) R_{sym}(q,n)}
  U_{sym}(q,-n) \nonumber\\& = \frac{1}{2  L}\sum_{q,q',n} <\rho_{sym}(q,n)\rho_{sym}(q',-n)>   \mbox{  }   U_{sym}(q,n)
  U_{sym}(q',-n)
  \end{align}
  This allows us to write down the interacting density-density correlations in momentum space
  \begin{align}
  <\rho_{sym}(q,n)\rho_{sym}(-q,-n)>\mbox{ }=\mbox{ } \frac{4 \pi  q (q v_F-i w_n)}{\beta  \left(4 \pi ^2 \left(q^2 v_F^2+w_n^2\right)- q^2 v_0^2\right)}
  \label{rhosymmom1}
 \end{align}
 and 
  \begin{align}
   <\rho_{sym}(q,n)\rho_{sym}(q,-n)>\mbox{   } = \mbox{   }
   -\frac{2 q^2 v_0}{\beta  \left(4 \pi ^2 \left(q^2 v_F^2+w_n^2\right)-q^2 v_0^2\right)}
   \label{rhosymmom2}
\end{align}
The correlations in space-time are calculated using
\begin{align}
 <T\rho_{sym}(x,t)\rho_{sym}(x',t')> \mbox{  } = \mbox{  }
      &\frac{1}{L}\sum_{q,n} e^{-iq (x-x') } \mbox{ } e^{w_n(t-t') }
      \mbox{        } <\rho_{sym}(q,n)\rho_{sym}(-q,-n)>  \nonumber\\& +       \frac{1}{L}\sum_{q,n} e^{-iq (x+x') } \mbox{ } e^{w_n(t-t') }
      \mbox{        } <\rho_{sym}(q,n)\rho_{sym}(q,-n)>
 \end{align}
 and we obtain the most singular part of the correlation function of the symmetric density fields
 \begin{align}
 <T\rho_{sym}(x,t)\rho_{sym}(x',t')> \mbox{  } = \mbox{  }\frac{\substack{v_{0} \left(csch^{2}\left(\frac{\pi  ((t-t') v_{h}+x+x')}{\beta  v_{h}}\right)+csch^{2}\left(\frac{\pi  (-(t-t') v_{h}+x+x')}{\beta  v_{h}}\right)\right)\\ + \mbox{ }2 \pi  \left((v_{h}-v_{F}) csch^{2}\left(\frac{\pi  ((t-t') v_{h}+x-x')}{\beta  v_{h}}\right)-(v_{F}+v_{h}) csch^{2}\left(\frac{\pi  ((t-t') v_{h}-x+x')}{\beta  v_{h}}\right)\right)}}{\substack{8 \pi  \beta^{2}  v_{h}^3}}
 \label{rsymrsymint}
 \end{align}
 where the holon velocity $v_{h}$ is defined by the relation $v_{h}^{2} = v_{F}^{2} - \frac{v_{0}^{2}}{4 \pi^{2}}$. This is the velocity at which the charge-carrying excitations (holons) propogate through the system. It's a fundamental parameter in describing the dynamics of one-dimensional systems with strong fermion-fermion interactions. In Luttinger liquids of fermions with spin, the charge excitations propagate with velocity $v_{h}$ while spin excitations (spinons) have a velocity equal to the Fermi velocity $v_{F}$. This spin-charge separation is a hallmark of Luttinger liquid physics. The holon velocity depends on the strength of the mutual interactions and it reduces to the Fermi velocity $v_{F}$ when $v_{0} = 0$. In the non-interacting limit ($v_{0} \rightarrow 0$) Eq.\ref{rsymrsymint} reduces to Eq.\ref{rsymrsym}.
 \section{Correlation function of the antisymmetric density fields}
 \label{antisymmetriccorrel}
 In this section we calculate the correlation function of the antisymmetric densities from the generating functional. The most singular part of the correlation function of the $\rho_{asy}$ fields will involve the bare reflection and transmission coefficients of the point-contact impurity. We obtain the saddle point of the $U^{'}_{asy}$ field by solving Eq.\ref{uasy}. In momentum and frequency space we have
 \begin{align}
 -i \beta \sum_{q'}  < \rho_{asy}(q,n)\rho_{asy}(q',-n) >_{0} U^{'}_{asy}(-q',n) - \frac{2 i}{v_{0}}( U_{asy}(-q,n) - U^{'}_{asy}(-q,n)) = 0
 \label{uasymomentum}
 \end{align}
 The Fourier transform to momentum-Matsubara frequency space of the non-interacting correlations is
 \begin{align}
 < \rho_{asy}(q,n)\rho_{asy}(q',-n) >_{0} \mbox{ }=\mbox{ } -\frac{(i q v_{F}) \delta _{q+q',0}}{(\pi  \beta  v_{F}) (w_{n}-i q v_{F})}-\frac{(r_{1}-1) w_{n} sgn(w_{n})}{\pi  \beta  L (q v_{F}+i w_{n}) (q' v_{F}-i w_{n})}
 \label{rhoasyqn}
 \end{align}
 Therefore we may write Eq.\ref{uasymomentum} in the form
 \begin{align}
 \frac{-i  (-1+r_{1}) w_{n} sgn(w_{n})}{\pi (q v_{F} + i w_{n})} U^{'}_{H,asy}(n) + \frac{i }{\pi v_{F}}\frac{i v_{F} q}{(w_{n}-i v_{F} q)}U^{'}_{asy}(q,n) - \frac{2 i}{v_{0}}(U_{asy}(-q,n)-U^{'}_{asy}(-q,n)) = 0
 \end{align}
 where we have introduced the Hilbert transform $U^{'}_{H,asy}(n)$ which is defined as \cite{Danny_Babu_2020,king2009hilbert}
 \begin{align}
 U^{'}_{H,asy}(n) = \frac{1}{L}\sum_{q'}\frac{ U^{'}_{asy}(q',n)}{(q' v_{F} + i w_{n})}
 \end{align}
 This allows us to obtain the saddle point in terms of the Hilbert transform
 \begin{align}
 U^{'}_{asy}(q,n) = \frac{\substack{\bigg((r_{1}-1) v_{0} w_{n} U^{'}_{H,asy}(n) sgn(w_{n}) (-q v_{0}+2 \pi  q v_{F}+2 i \pi  w_{n})\\-2 \pi  (q v_{F}+i w_{n}) (q v_{0} U_{asy}(-q,n)+2 \pi  (q v_{F}-i w_{n}) U_{asy}(q,n))\bigg)}}{q^2 \left(v_{0}^2-4 \pi ^2 v_{F}^2\right)-4 \pi ^2 w_{n}^2}
 \end{align}
 The generating functional for the $U_{asy}$ fields is
 \begin{align}
 Z[U_{asy}] = e^{  -\frac{1}{i v_{0}} \int_C dt \int dx \mbox{ } (U_{asy}(-x,t) - U^{'}_{asy}(-x,t)) U_{asy}(x,t)}
 \end{align}
 which in Fourier space is
 \begin{align}
 Z_{asy}[U] = e^{  -\frac{1}{ v_{0} \beta L } \sum_{q,n}
  (U_{asy}(q,n)-U^{'}_{asy}(q,n))
  U_{asy}(q,-n)}
  \end{align}
  Using the definition of the Hilbert transform we may write,
  \begin{align}
  U^{'}_{H,asy}(n) = -\sum_{q}\frac{2 \pi  (q v_{0} U_{asy}(-q,n)+2 \pi  (q v_{F}-i w_{n}) U_{asy}(q,n))}{L \left(q^2 \left(v_{0}^2-4 \pi ^2 v_{F}^2\right)-4 \pi ^2 w_{n}^2\right) \left(\frac{(r_{1}-1) (v_{0}+2 \pi  (v_{F}-v_{h}))}{2 (v_{0}+2 \pi  v_{F})}+1\right)}
  \end{align}
  \begin{figure}
\centering
 \includegraphics[scale=0.6]{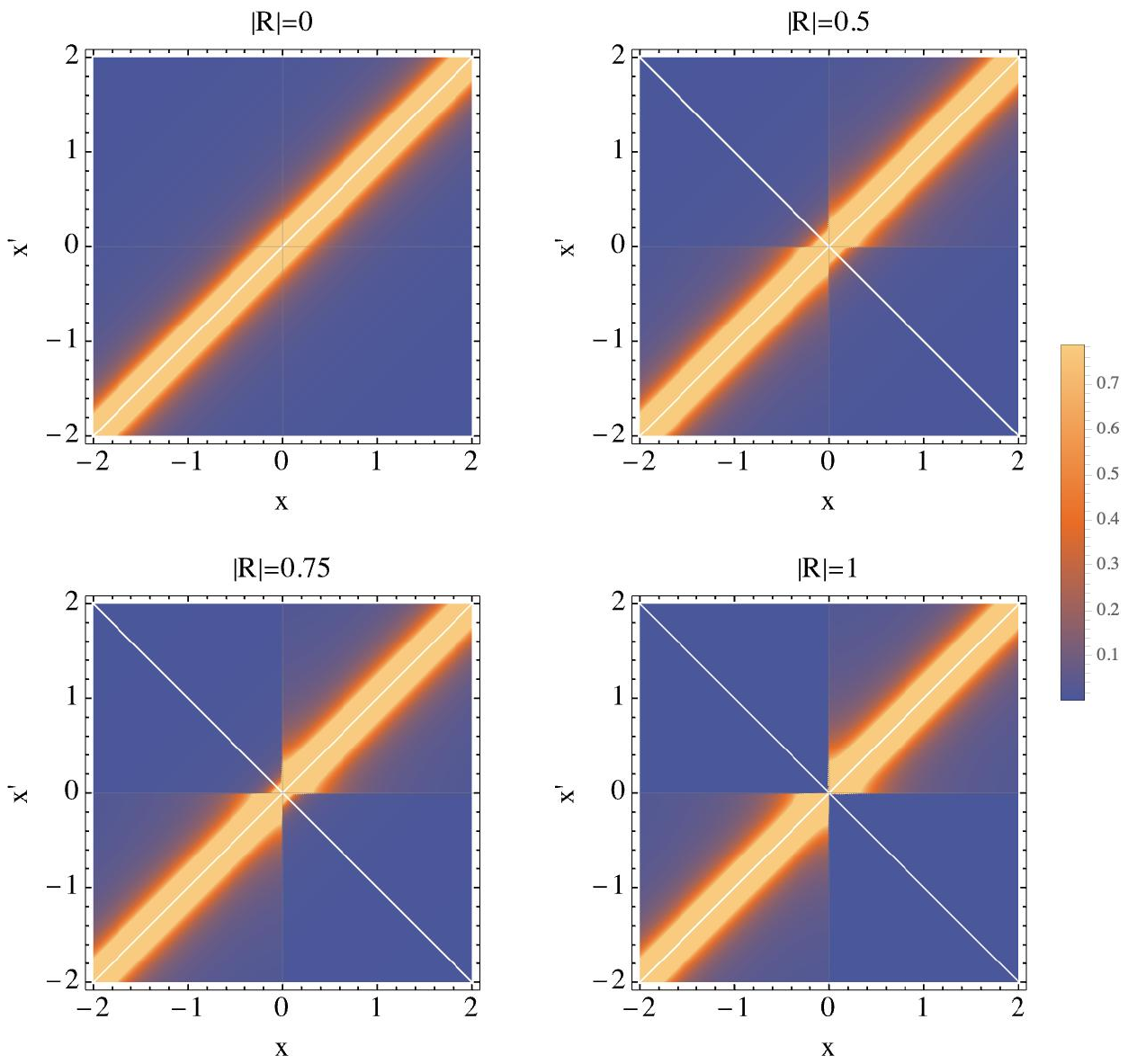}
 \caption{\small Equal time right mover density-density correlation function $|<\rho_{R}(x,t)\rho_{R}(x',t)>|$ as a function of spatial coordinates $x$ and $x'$ for different values of the reflection amplitude $|R|$ of the point-contact impurity for filling fraction $\nu = 1/3$, where $v_{0} = \frac{\pi  \left(2 \left(\nu ^2-1\right)\right) v_{F}}{\nu ^2+1}$. The correlation in the region near the origin with $x$ and $x'$ on opposite sides gets progressively reduced with increasing impurity reflection amplitude. $v_{F} \beta = 100$ for the plots.}
 \label{rhorhoplot}
 \end{figure}
 Hence we have,
 \begin{align}
\ln(&Z[U_{asy}]) = \nonumber\\&-\Sigma _{q,n}\frac{v_{0}  (2 \pi  q (q v_{F}+i w_{n}) U_{asy}(-q,n) U_{asy}(q,-n))}{(\beta  L v_{0}) \left(q^2 \left(v_{0}^2-4 \pi ^2 v_{F}^2\right)-4 \pi ^2 w_{n}^2\right)}-\Sigma _{q,n}\frac{q^2 v_{0}^{2}  U_{asy}(q,n) U_{asy}(q,-n)}{(\beta  L v_{0}) \left(q^2 \left(v_{0}^2-4 \pi ^2 v_{F}^2\right)-4 \pi ^2 w_{n}^2\right)}\nonumber\\
&+\Sigma _{p,q,n}\frac{\substack{(r_{1}-1) v_{0} w_{n} sgn(w_{n})  (q v_{0}-2 \pi  q v_{F}-2 i \pi  w_{n})  2 \pi  (-p v_{0} +2 \pi  (p v_{F}-i w_{n}))U_{asy}(p,n) U_{asy}(q,-n) }}{\substack{\left(\beta  L^2 v_{0}\right) \left(\left(q^2 \left(v_{0}^2-4 \pi ^2 v_{F}^2\right)-4 \pi ^2 w_{n}^2\right) \left(\left(p^2 \left(v_{0}^2-4 \pi ^2 v_{F}^2\right)-4 \pi ^2 w_{n}^2\right) \left(\frac{(r_{1}-1) (v_{0}+2 \pi  (v_{F}-v_{h}))}{2 (v_{0}+2 \pi  v_{F})}+1\right)\right)\right)}}
\end{align}
  This allows us to evaluate the most singular part of the correlation from the relation
  \begin{align}
&-\Sigma _{q,n}\frac{v_{0}  (2 \pi  q (q v_{F}+i w_{n}) U_{asy}(-q,n) U_{asy}(q,-n))}{(\beta  L v_{0}) \left(q^2 \left(v_{0}^2-4 \pi ^2 v_{F}^2\right)-4 \pi ^2 w_{n}^2\right)}-\Sigma _{q,n}\frac{q^2 v_{0}^{2}  U_{asy}(q,n) U_{asy}(q,-n)}{(\beta  L v_{0}) \left(q^2 \left(v_{0}^2-4 \pi ^2 v_{F}^2\right)-4 \pi ^2 w_{n}^2\right)}\nonumber\\
&+\Sigma _{p,q,n}\frac{\substack{(r_{1}-1) v_{0} w_{n} sgn(w_{n})  (q v_{0}-2 \pi  q v_{F}-2 i \pi  w_{n})  2 \pi  (-p v_{0} +2 \pi  (p v_{F}-i w_{n}))U_{asy}(p,n) U_{asy}(q,-n) }}{\substack{\left(\beta  L^2 v_{0}\right) \left(\left(q^2 \left(v_{0}^2-4 \pi ^2 v_{F}^2\right)-4 \pi ^2 w_{n}^2\right) \left(\left(p^2 \left(v_{0}^2-4 \pi ^2 v_{F}^2\right)-4 \pi ^2 w_{n}^2\right) \left(\frac{(r_{1}-1) (v_{0}+2 \pi  (v_{F}-v_{h}))}{2 (v_{0}+2 \pi  v_{F})}+1\right)\right)\right)}}\nonumber\\
&\mbox{ }=\mbox{ }\frac{1}{2 L}\sum_{p,q,n}\mbox{  }
  <\rho_{asy}(p,n)\rho_{asy}(q,-n)> \mbox{   }
 U_{asy}(p,n) U_{asy}(q,-n)
\end{align}
We obtain
\begin{align}
<T&\mbox{ }\rho_{asy}(x,t) \rho_{asy}(x',t')>\mbox{ }=\mbox{ }-\frac{2}{v_{0} \beta L} \Sigma _{q,n}e^{-i q(-x+x')}e^{w_{n}(t-t')}\frac{v_{0} (2 \pi  q (q v_{F}+i w_{n}) )}{ \left(q^2 \left(v_{0}^2-4 \pi ^2 v_{F}^2\right)-4 \pi ^2 w_{n}^2\right)}\nonumber\\
&-\frac{2}{v_{0} \beta L} \Sigma _{q,n}e^{-i q(x+x')}e^{w_{n}(t-t')}\frac{q^2 v_{0}^{2}}{\left(q^2 \left(v_{0}^2-4 \pi ^2 v_{F}^2\right)-4 \pi ^2 w_{n}^2\right)}\nonumber\\
&+\Sigma_{q,q',n}e^{-i q x} e^{-i q' x'} e^{w_{n}(t-t')} \frac{\substack{2 (r_{1}-1) v_{0} w_{n} sgn(w_{n}) (2 \pi  (-q v_{0}+2 \pi  (q v_{F}-i w_{n}))) (q' v_{0}-2 \pi  q' v_{F}-2 i \pi  w_{n})}}{\substack{(\beta  L^{2} v_{0}) \left(q^{'2} \left(-4 \pi ^2 v_{h}^2\right)-4 \pi ^2 w_{n}^2\right) \left(\left(q^2 \left(-4 \pi ^2 v_{h}^2\right)-4 \pi ^2 w_{n}^2\right) \left(\frac{(r_{1}-1) (v_{0}+2 \pi  (v_{F}-v_{h}))}{2 (v_{0}+2 \pi  v_{F})}+1\right)\right)}}
\label{rhoasyrhoasyft}
\end{align}
The summations are evaluated analytically with the use of Mathematica software \cite{Mathematica}. The Fourier transform from momentum to spatial coordinates is performed first followed by the summation over the bosonic Matsubara frequencies. The correlations in space-time are obtained as
\begin{align}
<T&\mbox{ }\rho_{asy}(x,t) \rho_{asy}(x',t')>\mbox{ }=\mbox{ } \nonumber\\&-\frac{\substack{v_{0} \left(csch^2\left(\frac{\pi  (\tau  v_{h}+x+x')}{\beta  v_{h}}\right)+csch^2\left(\frac{\pi  (-\tau  v_{h}+x+x')}{\beta  v_{h}}\right)\right)}}{\substack{8 \pi  \beta ^2 v_{h}^3}}-\frac{\substack{(v_{F}-v_{h}) csch^2\left(\frac{\pi  (\tau  v_{h}+x-x')}{\beta  v_{h}}\right)+(v_{F}+v_{h}) csch^2\left(\frac{\pi  (\tau  v_{h}-x+x')}{\beta  v_{h}}\right)}}{\substack{4 \beta ^2 v_{h}^3}}\nonumber\\
& -\frac{\substack{\theta(x x') \left((r_{1}-1) (v_{0}+2 \pi  v_{F}) (v_{0}+2 \pi  (v_{h}-v_{F})) (2 \pi  (v_{F}+v_{h})-v_{0}) \left(csch^2\left(\frac{\pi  (\tau  v_{h}+x+x')}{\beta  v_{h}}\right)+csch^2\left(\frac{\pi  (-\tau  v_{h}+x+x')}{\beta  v_{h}}\right)\right)\right)}}{\substack{16 \pi ^2 \beta ^2 v_{h}^4 ((r_{1}+1) v_{0}+2 \pi  (r_{1} v_{F}-r_{1} v_{h}+v_{F}+v_{h}))}}\nonumber\\
&+ \theta(-x x') \frac{\substack{(r_{1}-1) (v_{0}+2 \pi  v_{F}) \left(-(v_{0}-2 \pi  (v_{F}+v_{h}))^2 csch^2\left(\frac{\pi  (-\tau  v_{h}+x-x')}{\beta  v_{h}}\right)-(v_{0}+2 \pi  (v_{h}-v_{F}))^2 csch^2\left(\frac{\pi  (-\tau  v_{h}-x+x')}{\beta  v_{h}}\right)\right)}}{\substack{16 \pi ^2 \beta ^2 v_{h}^4 ((r_{1}+1) v_{0}+2 \pi  (r_{1} v_{F}-r_{1} v_{h}+v_{F}+v_{h}))}}
\label{rhoasyrhoasyst}
\end{align}  
where $\tau = t-t'$ and $v_{h}^{2} = v_{F}^{2} - \frac{v_{0}^{2}}{4 \pi^{2}}$. We have defined $r_1 = \left(1
  -\frac{32 \Gamma ^2 v_F^2}{\left(\Gamma ^2+4 v_F^2\right)^2}
\right)$. The point-contact tunneling amplitude is related to the bare reflection and transmission amplitudes of the point-contact impurity \cite{Babu_2022}.
\section{Results: Density-density correlation function}
\label{finresults}
The density-density correlation functions of the right and left moving fermions can be written compactly in terms of the correlations of the symmetric and antisymmetric densities in the following manner,
 \begin{align}
 <T\mbox{ }\rho_{\chi}(x,t)\rho_{\chi'}(x',t')>\mbox{ }=\mbox{ }\frac{1}{4}(<T \mbox{ } \rho_{sym}(\chi \mbox{ } x,t) \rho_{sym}(\chi'\mbox{ } x',t') > + \chi \chi' <T \mbox{ } \rho_{asy}(\chi \mbox{ }x,t) \rho_{asy}(\chi' \mbox{ }x',t') >)
 \label{rhorhoresult}
 \end{align}
 where $\chi$ and $\chi'$ can take values $\pm 1$ for $R$ and $L$ movers respectively. The analytical expressions for the correlations we obtain are nothing but the resummation of the most singular parts of the RPA terms in powers of the mutual interaction. In Fig.\ref{rhorhoplot} the equal time right mover density correlations are plotted for various values of the impurity reflection amplitude $|R|$. The point-contact is modelled as an isolated impurity and its tunneling amplitude is related to the reflection amplitude by $|R| = |-i \frac{4 v_{F} \Gamma}{\Gamma^{2} + 4 v_{F}^{2}}|$ \cite{Babu_2022}. With increasing impurity reflection amplitude the correlations in the region near the origin of the figure with $x$ and $x'$ on opposite sides of the impurity is reduced as a result of the localised impurity backscattering. Our result for the density correlations is not the full story, it captures only the singular behaviour of the correlations. The Gaussian approximation becomes exact only for a homogeneous system since the higher order moments of the density are zero in this case.
\section{Comparison with perturbation theory}
\label{pertcomparison}
In this section we compare the results we have obtained using the generating functional method with those obtained using standard fermionic perturbation theory. We expand the most singular parts of the DDCF we have obtained in powers of the interaction parameter $v_{0}$ and perform a term by term comparison with S-matrix perturbation expansion of the DDCF retaining only the most singular terms. We explicitly show upto $O(v_{0}^{2})$ that each term in the perturbation expansion matches with the S-matrix expansion. Our idea of the most singular truncation of the RPA generating functional gives us the correlation functions resummed to include only the most singular parts of the RPA terms. Perturbation expansion of our result in powers of $v_{0}$ will match with the corresponding series obtained by standard perturbation theory so long as one retains only the most singular terms.\\
\subsection{Perturbative comparison of $<\rho_{sym}\rho_{sym}>$}
The density-density correlations in presence of interactions is written in terms of the non-interacting ones in the following manner,
\begin{align}
<T \mbox{ } \rho_{sym}(x_{1},t_{1}) \rho_{sym}(x_{2},t_{2}) > \mbox{ }=\mbox{ } \frac{<T \mbox{ }S\mbox{ } \rho_{sym}(x_{1},t_{1}) \rho_{sym}(x_{2},t_{2}) >_{0}}{<T\mbox{ }S>_{0}}
\end{align}
  where $S\mbox{ }=\mbox{ }e^{- i \int_{c} dt H_{int}}$ and $H_{int}$ is given by Eq.\ref{hint}. The perturbaton expansion in powers of $v_{0}$ gives
  \begin{align}
  <T \mbox{ } &\rho_{sym}(x_{1},t_{1}) \rho_{sym}(x_{2},t_{2}) > \mbox{ }=\mbox{ } \frac{<T \mbox{ }S\mbox{ } \rho_{sym}(x_{1},t_{1}) \rho_{sym}(x_{2},t_{2}) >_{0}}{<T\mbox{ }S>_{0}}\nonumber\\
  &=\mbox{ }<T \mbox{ } \rho_{sym}(x_{1},t_{1}) \rho_{sym}(x_{2},t_{2}) >_{0} - i \int_{C} dt <T\mbox{ }H_{int}(t) \rho_{sym}(x_{1},t_{1}) \rho_{sym}(x_{2},t_{2})>_{0,c} \nonumber\\&\mbox{ }\mbox{ }\mbox{ }\mbox{ }\mbox{ }\mbox{ }+ \frac{(-i)^{2}}{2} \int_{C} dt \int_{C} dt' <T \mbox{ }H_{int}(t) H_{int}(t')  \rho_{sym}(x_{1},t_{1}) \rho_{sym}(x_{2},t_{2}) >_{0,c} + ...........
  \label{pertexprhosym}
  \end{align}
  where $<\mbox{ }\mbox{ }>_{0,c}$ denotes the non-interacting connected correlation functions. We retain only the most singular terms and use Eqs.\ref{rhosymasy0}, \ref{rhoasysym0} and \ref{hintsymasy} and express the first order term in the perturbation series in the form (note that $<\rho_{sym} \rho_{asy} >_{0} = 0$),
  \begin{align}
  <T \mbox{ } &\rho_{sym}(x_{1},t_{1}) \rho_{sym}(x_{2},t_{2}) >^{(1)} \mbox{ }\nonumber\\&=\mbox{ } - i \frac{v_{0}}{4} \int_{C} dt \int dx \bigg(<T \mbox{ } \rho_{sym}(x,t)\rho_{sym}(x_{1},t_{1})>_{0}<T \mbox{ } \rho_{sym}(-x,t)\rho_{sym}(x_{2},t_{2})>_{0}\nonumber \\
  &\mbox{ }\mbox{ }\mbox{ }\mbox{ }\mbox{ }\mbox{ }+ <T \mbox{ } \rho_{sym}(x,t)\rho_{sym}(x_{2},t_{2})>_{0}<T \mbox{ } \rho_{sym}(-x,t)\rho_{sym}(x_{1},t_{1})>_{0}\bigg)
  \end{align}
  It is more convenient to work in momentum and frequency space,
  \begin{align}
  <&\rho_{sym}(q',n)\rho_{sym}(q'',-n) >^{(1)} \mbox{ }\nonumber\\&=\mbox{ } -\frac{v_{0} \beta}{4} \bigg(\sum_{q} <\rho_{sym}(q,-n)\rho_{sym}(q',n)>_{0} <\rho_{sym}(q,n)\rho_{sym}(q'',-n)>_{0} \nonumber \\
  &\mbox{ }\mbox{ }\mbox{ }\mbox{ }\mbox{ }\mbox{ }+ \sum_{q} <\rho_{sym}(q,n)\rho_{sym}(q'',-n)>_{0}<\rho_{sym}(q,-n)\rho_{sym}(q',n)>_{0}\bigg)
  \end{align}
  Making use of  $<\rho_{sym}(q,n)\rho_{sym}(q',-n)>_{0} = -\frac{1}{\pi \beta v_{F}}\frac{  i v_F q}{(w_{n}-i v_{F} q)} \delta_{q+q',0}$ we get,
  \begin{align}
  <\rho_{sym}(q',n)\rho_{sym}(q'',-n) >^{(1)} = -\frac{q^{'2} v_{0}}{2 \left(\pi ^2 \beta  \left(q^{'2} v_{F}^2+w_{n}^2\right)\right)}\mbox{ }\delta_{q',q''}
  \label{rhosympert1}
  \end{align}
  It is easy to see that the first order term in the expansion of Eq.\ref{rhosymmom2} in powers of $v_{0}$ is equal to the expression in Eq.\ref{rhosympert1}. The second order term in the expansion in Eq.\ref{pertexprhosym} is written as
  \begin{align}
  <T \mbox{ }& \rho_{sym}(x_{1},t_{1}) \rho_{sym}(x_{2},t_{2})>^{(2)} \mbox{ }\nonumber\\&=\mbox{ } \substack{\frac{(-i v_{0})^{2}}{32} \int_{C} dt \int_{C} dt' \int dx \int dx' <T \mbox{ } \rho_{sym}(x,t)\rho_{sym}(-x,t) \rho_{sym}(x',t') \rho_{sym}(-x',t') \rho_{sym}(x_{1},t_{1}) \rho_{sym}(x_{2},t_{2}) >_{0,c}}
  \end{align}
  Evaluating this term in the momentum and frequency space (refer to \hyperref[AppendixA]{Appendix A}) while retaining only the most singular parts we get,
  \begin{align}
  <\rho_{sym}(q',n)\rho_{sym}(q'',-n) >^{(2)}\mbox{ }=\mbox{ }\frac{q^{'3} v_{0}^2}{4 \pi^3 \beta  (q' v_{F}-i w_{n}) (q' v_{F}+i w_{n})^2}\mbox{ }\delta_{q'+q'',0}
  \end{align}
  This is precisely equal to the second order term we get on expanding Eq.\ref{rhosymmom1} in powers of $v_{0}$.
  \subsection{Perturbative comparison of $<\rho_{asy}\rho_{asy}>$}
  Similar to the previous section the standard perturbation expansion of the $<\rho_{asy}\rho_{asy}>$ correlation function in powers of $v_{0}$ is,
  \begin{align}
   <T \mbox{ } &\rho_{asy}(x_{1},t_{1}) \rho_{asy}(x_{2},t_{2}) > \mbox{ }=\mbox{ } \frac{<T \mbox{ }S\mbox{ } \rho_{asy}(x_{1},t_{1}) \rho_{asy}(x_{2},t_{2}) >_{0}}{<T\mbox{ }S>_{0}}\nonumber\\
  &=\mbox{ }<T \mbox{ } \rho_{asy}(x_{1},t_{1}) \rho_{asy}(x_{2},t_{2}) >_{0} - i \int_{C} dt <T\mbox{ }H_{int}(t) \rho_{asy}(x_{1},t_{1}) \rho_{asy}(x_{2},t_{2})>_{0,c} \nonumber\\&\mbox{ }\mbox{ }\mbox{ }\mbox{ }\mbox{ }\mbox{ }+ \frac{(-i)^{2}}{2} \int_{C} dt \int_{C} dt' <T \mbox{ }H_{int}(t) H_{int}(t')  \rho_{asy}(x_{1},t_{1}) \rho_{asy}(x_{2},t_{2}) >_{0,c} + ...........
  \end{align}
  Retaining only the most singular parts, the first order term can be written as (note that $<\rho_{sym} \rho_{asy} >_{0} = 0$),
  \begin{align}
  <T \mbox{ } &\rho_{asy}(x_{1},t_{1}) \rho_{asy}(x_{2},t_{2}) >^{(1)} \mbox{ }\nonumber\\&=\mbox{ }\frac{i v_{0}}{4} \int_{c} dt \int dx \bigg(<T\mbox{ }\rho_{asy}(x,t)\rho_{asy}(x_{1},t_{1})>_{0} <T\mbox{ }\rho_{asy}(-x,t)\rho_{asy}(x_{2},t_{2})>_{0}\nonumber \\&\mbox{ }\mbox{ }\mbox{ }\mbox{ }\mbox{ }\mbox{ }+ <T\mbox{ }\rho_{asy}(x,t)\rho_{asy}(x_{2},t_{2})>_{0} <T\mbox{ } \rho_{asy}(-x,t)\rho_{asy}(x_{1},t_{1})>_{0}\bigg)
  \end{align}
  In frequency and momentum space this is
  \begin{align}
  <  &\rho_{asy}(q',n) \rho_{asy}(q'',-n) >^{(1)}\mbox{ }\nonumber\\&=\mbox{ }\frac{i v_{0}}{4} (-i \beta)\bigg(\sum_{q} <\rho_{asy}(q,-n) \rho_{asy}(q',n)>_{0}<\rho_{asy}(q,n) \rho_{asy}(q'',-n)>_{0} \nonumber \\&\mbox{ }\mbox{ }\mbox{ }\mbox{ }\mbox{ }\mbox{ }+ \sum_{q}<\rho_{asy}(q,n) \rho_{asy}(q'',-n)>_{0}<\rho_{asy}(q,-n) \rho_{asy}(q',n)>_{0}\bigg)
  \end{align}
  Using Eq.\ref{rhoasyqn} we evaluate this term and obtain
  \begin{align}
  <  &\rho_{asy}(q',n) \rho_{asy}(q'',-n) >^{(1)}\mbox{ }\nonumber\\&=\mbox{ } v_{0} \bigg(\frac{\substack{(r_{1}-1) w_{n} sgn(w_{n}) \left(q v_{F}^2 (q' (2 (r_{1}+1) v_{F}-r_{1} v_{h}+v_{h})+i (r_{1}+1) w_{n})+w_{n} \left((r_{1}-1) v_{h} w_{n}-i q' (r_{1}+1) v_{F}^2\right)\right)}}{\pi ^2 \beta  L \left(q^2 v_{h}^2+w_{n}^2\right) \left(q^{'2} v_{h}^2+w_{n}^2\right) (r_{1} v_{F}-r_{1} v_{h}+v_{F}+v_{h})^2}\nonumber \\&\mbox{ }\mbox{ }\mbox{ }\mbox{ }\mbox{ }\mbox{ }+\frac{q^2 \delta _{q,q'}}{2 \pi ^2 \beta  \left(q^2 v_{F}^2+w_{n}^2\right)}\bigg)
  \end{align}
This is equal to the first order term obtained upon expanding the $<  \rho_{asy}(q',n) \rho_{asy}(q'',-n) >$ correlation function that can be inferred from the rhs of Eq.\ref{rhoasyrhoasyft}. The second order term in the expansion is
\begin{align}
<&T \mbox{ } \rho_{asy}(x_{1},t_{1}) \rho_{asy}(x_{2},t_{2}) >^{(2)}\mbox{ }\nonumber\\&=\mbox{ }\substack{\frac{(-i v_{0})^{2}}{32} \int_{C}dt \int_{C} dt' \int dx \int dx' <T \mbox{ }\rho_{asy}(x,t)\rho_{asy}(-x,t) \rho_{asy}(x',t') \rho_{asy}(-x',t') \rho_{asy}(x_{1},t_{1}) \rho_{asy}(x_{2},t_{2})>_{0,c}}
\end{align}
Retaining the most singular parts and evaluating this term (refer to \hyperref[AppendixB]{Appendix B}) in momentum and frequency space we get
\begin{align}
<  &\rho_{asy}(q',n) \rho_{asy}(q'',-n) >^{(2)}\mbox{ }\nonumber\\&=\mbox{ }v_{0}^2 \bigg(-\frac{\substack{(r_{1}-1) w_{n} sgn(w_{n})} \bigg(\substack{q' \left(q'' v_{F} \left(\left(r_{1}^2-1\right) v_{F} v_{h}+(r_{1}+1)^2 v_{F}^2-(r_{1}-1)^2 v_{h}^2\right)+i (r_{1}-1) v_{h} w_{n} (2 (r_{1}+1) v_{F}-r_{1} v_{h}+v_{h})\right)\\+(r_{1}-1) v_{h} w_{n} ((r_{1}+1) w_{n}-i q'' (2 (r_{1}+1) v_{F}-r_{1} v_{h}+v_{h}))\bigg)}}{\substack{2 \pi ^3 \beta  L \left(q'^2 v_{h}^2+w_{n}^2\right) \left(q''^{2} v_{h}^2+w_{n}^2\right) (r_{1} v_{F}-r_{1} v_{h}+v_{F}+v_{h})^3}}\nonumber \\&\mbox{ }\mbox{ }\mbox{ }\mbox{ }\mbox{ }\mbox{ }+\substack{\frac{q'^3 \delta _{q',-q''}}{4 \pi^3 \beta  (q' v_{F}-i w_{n}) (q' v_{F}+i w_{n})^2}}\bigg)
\end{align}
 This is exactly equal to the second order term obtained upon perturbative expansion of the $<  \rho_{asy}(q',n) \rho_{asy}(q'',-n) >$ correlation function (present in the rhs of Eq.\ref{rhoasyrhoasyft}) in powers of $v_{0}$. 
 \section{Two-terminal current in response to a difference in potential between the edges}
 \label{linearcurrent}
 In this section, the obtained DDCF is used to evaluate the two-terminal conductance (source-drain conductance) both in the absence and presence of the point-contact. The right movers and left movers are separated with a different potential for the right and left moving edges. We consider that the difference in chemical potentials is $\mu_{L} - \mu_{R} = -\mu_{R} = e V$, that is a bias is applied to the right movers and the chemical potential of the left movers is taken to be zero. This means we have
 \begin{align}
H_{bias} = -e V \int dy \rho_{R}(y,t)
\end{align}
The current is calculated using the definition \cite{giamarchi2003quantum},
\begin{align}
 j_{e}(x,t)\mbox{ }=\mbox{ }e v_{h} \nu (\Pi_{R}(x,t) - \Pi_{L}(x,t))\mbox{ }=\mbox{ }e v_{h} (\rho_{R}(x,t) - \rho_{L}(x,t))
 \end{align}
 where $\Pi_{\chi}(x,t) = \rho_{\chi}(x,t)/\nu$ is the momentum conjugate to the chiral $\phi_{\chi}(x,t)$ fields (defined by $\rho_{\chi}(x,t) = \frac{1}{2\pi}\partial_{x}\phi_{\chi}(x,t)$) for fractional quantum Hall edge states with $\nu = 1/m$ where $m$ is an odd integer \cite{doi:https://doi.org/10.1002/9783527617258.ch4} and $\chi = R$ or $L$. The average current is evaluated using
 \begin{align}
 <j_{e}(x,t)>\mbox{ }=\mbox{ }e v_{h}\frac{<T \mbox{ }S (\rho_{R}(x,t) - \rho_{L}(x,t))>_{V_{b}=0}}{<T S>_{V_{b}=0}}
 \end{align}
 where $S = e^{-i \int_{C} dt H_{bias}}$. The current in linear response to the bias maybe computed directly
 \begin{align}
 <j_{e}(x,t)>\mbox{ }=\mbox{ }e v_{h} (i e V) \int_{-\infty}^{\infty}dy \int_{0}^{-i \beta}dt_{1}\left(<\rho_{R}(y,t_{1}) \rho_{R}(x,t)> - <\rho_{R}(y,t_{1}) \rho_{L}(x,t)>\right)
 \end{align}
 In the absence of the point-contact ($\Gamma=0$) we use Eq.\ref{rhorhoresult} with $r_{1}=1$ and evaluate the current to obtain
 \begin{align}
 <j_{e}>\mbox{ }=\mbox{ }\frac{e^{2}}{h} \frac{v_{0} + 2 \pi v_{F}}{2 \pi v_{h}} V
 \end{align}
 The interaction term in our Hamiltonian is identical to the $g2$ term in the g-ology formalism and the filling fraction $\nu$ is identified with the Luttinger parameter \cite{giamarchi2003quantum}, hence the interaction parameter $v_{0}$ of the Luttinger liquid can related to the FQHE bulk filling fraction through the equation $v_{0} = \frac{\pi  \left(2 \left(\nu ^2-1\right)\right) v_{F}}{\nu ^2+1}$. This leads to the correct formula for the Hall conductance in the absence of the point-contact
 \begin{align}
 G = \nu \frac{e^{2}}{h}
 \end{align}
 In the presence of a point-contact, there appears an additional term quantifying the loss of current due to backscattering at the point-contact. The linear response conductance in this case is given by
 \begin{align}
G\mbox{ }=\mbox{ }\frac{e^2 \nu  \left(\frac{r_{1}-1}{\nu +(\nu -1) r_{1}+1}+1\right)}{h}
\label{Gsd}
\end{align} 
This expression has a backscattering contribution to the source-drain conductance. This is an approximate result valid at zero temperature, obtained using only the most singular part of the density correlations which are important in the limit the time interval is small $t-t_{1} \rightarrow 0$ (or high frequency limit in Fourier space). It is only this regime that is captured by Eq.\ref{Gsd}. The out of equilibrium conductance as a function of voltage and temperature is described by universal scaling functions which the present approach doesn't provide. A nonequilibrium extension of the present work which includes the calculation of the most-singular parts of the density-density correlation functions in presence of a voltage bias can be used to calculate the average current in the nonequilibrium system.
\begin{figure}
\centering
 \includegraphics[scale=0.7]{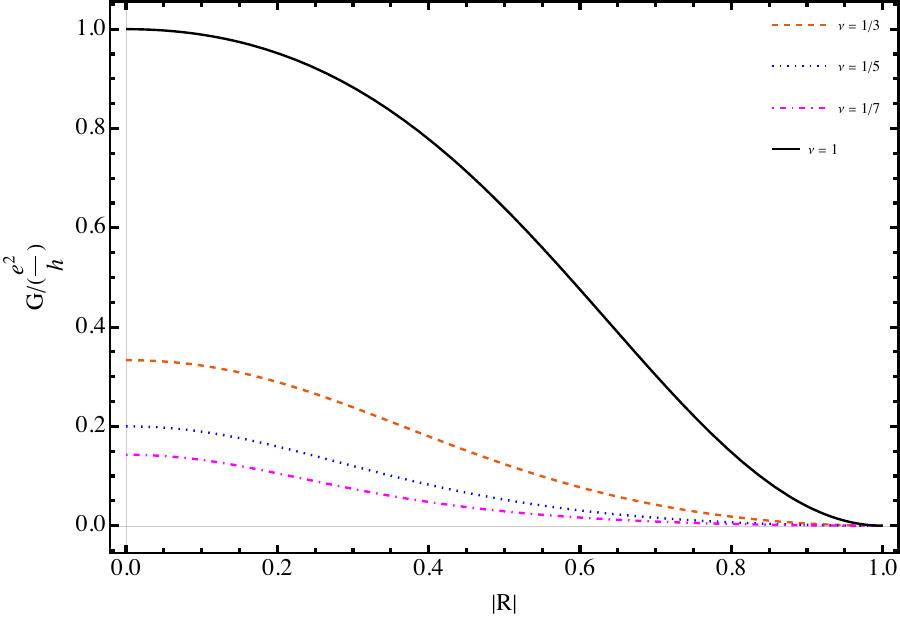}
 \caption{\small The linear two-terminal conductance in units of $\frac{e^{2}}{h}$ for edge states with different values of bulk filling fraction $\nu$ in the Laughlin series as a function of the reflection amplitude $|R|$ is shown. The solid black line indicates the non-interacting case $\nu=1$, which is the well known Landauer conductance. The backscattering (tunneling) contribution to the conductance is enhanced for successive filling fractions in the Laughlin series which results in a suppression of the source-drain conductance. When $|R| = 0$ the conductance $G = \nu \frac{e^{2}}{h}$. When $|R| = 1$ the conductance vanishes identically for all cases.}
 \label{lincond}
 \end{figure}
\subsection{Limiting case checks}
\textbf{No interactions:} In the absence of interactions $\nu=1$ the conductance reduces to
\begin{align}
G = \frac{e^{2}}{h}|T|^{2}
\end{align}
which is the Landauer's formula \cite{landauer} for conductance. Here we have used $r_1 = \left(1
  -\frac{32 \Gamma ^2 v_F^2}{\left(\Gamma ^2+4 v_F^2\right)^2}
\right)$ and the relation between the tunneling amplitude $\Gamma$ and the bare reflection and transmission amplitudes, $|R|$ and $|T|$ respectively \cite{Babu_2022},
\begin{align}
  |\frac{\Gamma}{v_{F}}| = \frac{2|R|}{1+|T|}
  \end{align}
  \textbf{No impurity:} In the absence of the point-contact we set $r_{1} = 1$ and we get back $G = \nu \frac{e^{2}}{h}$\\
  \mbox{ }\\
  \textbf{Half-line limit:} In the case of a half-line $|R|=1$ and we get
  \begin{align}
    G = 0 
  \end{align}   
  irrespective of the interactions. In Fig.\ref{lincond} the linear conductance in units of the conductance quantum is plotted as a function of the reflection amplitude of the point-contact impurity. The backscattering (tunneling) contribution to the conductance is enhanced for successive filling fractions in the Laughlin series and the source-drain conductance gets suppressed more as we progress along the Laughlin series of filling fractions. The domain of applicability of our method can be understood in the following manner: Our result for the linear conductance in presence of the point-contact is obtained with an exact non-perturbative treatment of the point-contact and mutual interparticle interactions, but it does not provide any information about the temperature dependence of the conductance. The most singular part of the correlations alone is not enough to obtain the full scaling functions of the tunneling current and conductance previously obtained in \cite{PhysRevB.52.8934} using a thermodynamic Bethe ansatz procedure for a certain range of filling fractions of the fractional quantum Hall bulk. 
 \section{Conclusions}
 \label{conclusions}
 In this work, the most singular part of the density-density correlation functions (DDCF) of chiral Luttinger liquids at the edges of fractional quantum Hall systems coupled through an interaction mediated by the bulk in presence of a point-contact impurity has been rigorously derived and shown to be expressible in terms of simple functions of positions and times. The expressions for the DDCF have simple second order poles and involve the scale independent bare reflection and transmission amplitudes of a single fermion in the presence of a localised impurity potential. Our results are compared with standard perturbation theory and are shown to match term by term. In the absence of the point-contact (homogeneous case) the Gaussian approximation is exact in the RPA sense and we obtain the standard expression for the Hall conductance. In the presence of the point-contact the Gaussian approximation corresponds to the most singular truncation of the cumulant expansion and this allows us to compute the correction to the Hall conductance due to the tunneling at the point-contact. Moreover,the DDCF result obtained in this work is used in a novel bosonization method to calculate the tunneling density of states (TDOS) for fractional quantum Hall edges coupled through a point-contact. The correct TDOS exponents for tunneling at the point-contact is obtained both in the electron tunneling and quasiparticle tunneling cases in \cite{babu2023tunneling} and is a crucial validation of our results. For the inhomogeneous system under consideration, the idea is to neglect all higher order (connected) moments of the density operator beyond the second order. The reason being that all odd moments of the density vanish identically whereas all the higher order even moments are less singular than the second moment as shown in \cite{Danny_Babu_2020}. This result is a crucial input to unconventional bosonization techniques \cite{babu2023unconventional,doi:10.1142/S0217751X18501749,das2019nonchiral} that retrieve the most singular part of the two-point correlations for arbitrary impurity strengths and interactions. On the other hand conventional bosonization methods are only able to obtain analytical expressions for the Green's functions in the homogeneous limit ($|R|=0$) or the half-line limit ($|R| = 1$) where higher order moments of the density beyond the second moment vanish identically. The question arises as to which physical quantity depends on the most singular part of the space-time correlation functions ? If we look at the local (single particle) density of states at the (spatial) origin i.e. $x_{1} = x_{2} = 0$, the Green's function is a function of the time interval $G(t_{1}-t_{2})$. This diverges as $t_{1} \rightarrow t_{2}$. This is the most singular part of this correlation function in the sense of the present manuscript. In frequency space we denote this as $D(\omega)$, where $\omega$ is the Fourier transform of $t_{1}-t_{2}$. This means that $t_{1}-t_{2} \rightarrow 0$ corresponds to $D(\omega \rightarrow \infty)$. Thus our formalism captures the high frequency limit of the dynamical density of states exactly. Similarly when we apply a bias to the system, a current flows in the system and this current is related to the equal space-equal-time 
( $ j = lim_{x_2 \rightarrow x_1} lim_{t_2 \rightarrow t_1}v_F ( <\psi^{\dagger}_R(x_1,t_1)\psi_R(x_2,t_2)> - <\psi^{\dagger}_L(x_1,t_1)\psi_L(x_2,t_2)>) $) limit of the correlations which is captured by our method. However, the study of non-equilibrium transport through the point-contact in the presence of a voltage bias requires the computation of the non-equilibrium density-density correlation functions, which would be an interesting and non-trivial extension of the present work because the interparticle interactions, impurity backscattering and voltage bias have to be treated non-perturbatively on an equal footing and this is still an open problem and is out of the scope of the present paper but will be dealt with in a future publication. 
   \section*{Acknowledgements}
  N.D.Babu acknowledges support through a doctoral fellowship provided by the Ministry of Education, Government of India.\\
  \mbox{ }\\
  
  \section*{APPENDIX  A: Perturbative comparison of second order ($O(v_{0}^{2})$) term for $<\rho_{sym}\rho_{sym}>$}
\label{AppendixA}
\setcounter{equation}{0}
\renewcommand{\theequation}{A.\arabic{equation}}

The second order term in the expansion in Eq.\ref{pertexprhosym} is
  \begin{align}
  <T \mbox{ }& \rho_{sym}(x_{1},t_{1}) \rho_{sym}(x_{2},t_{2})>^{(2)} \mbox{ }\nonumber\\&=\mbox{ } \substack{\frac{(-i v_{0})^{2}}{32} \int_{C} dt \int_{C} dt' \int dx \int dx' <T \mbox{ } \rho_{sym}(x,t)\rho_{sym}(-x,t) \rho_{sym}(x',t') \rho_{sym}(-x',t') \rho_{sym}(x_{1},t_{1}) \rho_{sym}(x_{2},t_{2}) >_{0,c}}
  \end{align}

Retaining only the most singular parts involves making the Gaussian approximation \cite{Danny_Babu_2020} which means that Wick's theorem is applicable at the level of pairs of fermions. This means we can write
\begin{align}
&<T \mbox{ } \rho_{sym}(x_{1},t_{1}) \rho_{sym}(x_{2},t_{2})>^{(2)} \mbox{ }\nonumber\\&=\mbox{ } \substack{\frac{(-i v_{0})^{2}}{32} \int_{C} dt \int_{C} dt' \int dx \int dx' <T \mbox{ } \rho_{sym}(x,t)\rho_{sym}(-x,t) \rho_{sym}(x',t') \rho_{sym}(-x',t') \rho_{sym}(x_{1},t_{1}) \rho_{sym}(x_{2},t_{2}) >_{0,c}}\nonumber\\
& = \substack{\frac{(-i v_{0})^{2}}{32} \int_{C} dt \int_{C} dt' \int dx \int dx' \bigg( <T\mbox{ }\rho_{sym}(x,t) \rho_{sym}(x_{1},t_{1})>_{0} <T\mbox{ }\rho_{sym}(x',t') \rho_{sym}(x_{2},t_{2})>_{0} <T\mbox{ }\rho_{sym}(-x,t) \rho_{sym}(-x',t')>_{0}}\nonumber\\&\mbox{ }\mbox{ }\substack{+ \mbox{ }<T\mbox{ }\rho_{sym}(x,t) \rho_{sym}(x_{2},t_{2})>_{0} <T\mbox{ }\rho_{sym}(x',t') \rho_{sym}(x_{1},t_{1})>_{0} <T\mbox{ }\rho_{sym}(-x,t) \rho_{sym}(-x',t')>_{0}}\nonumber\\& \mbox{ }\mbox{ }\substack{+ \mbox{ }<T\mbox{ }\rho_{sym}(-x,t) \rho_{sym}(x_{1},t_{1})>_{0} <T\mbox{ }\rho_{sym}(x',t') \rho_{sym}(x_{2},t_{2})>_{0} <T\mbox{ }\rho_{sym}(x,t) \rho_{sym}(-x',t')>_{0} }\nonumber\\& \mbox{ }\mbox{ }\substack{+ \mbox{ }<T\mbox{ }\rho_{sym}(-x,t) \rho_{sym}(x_{2},t_{2})>_{0} <T\mbox{ }\rho_{sym}(x',t') \rho_{sym}(x_{1},t_{1})>_{0} <T\mbox{ }\rho_{sym}(x,t) \rho_{sym}(-x',t')>_{0}}\nonumber\\& \mbox{ }\mbox{ }\substack{+ \mbox{ }<T\mbox{ }\rho_{sym}(-x',t') \rho_{sym}(x_{1},t_{1})>_{0} <T\mbox{ }\rho_{sym}(x,t) \rho_{sym}(x_{2},t_{2})>_{0} <T\mbox{ }\rho_{sym}(-x,t) \rho_{sym}(x',t')>_{0}}\nonumber\\& \mbox{ }\mbox{ }\substack{+\mbox{ }<T\mbox{ }\rho_{sym}(-x',t') \rho_{sym}(x_{2},t_{2})>_{0} <T\mbox{ }\rho_{sym}(x,t) \rho_{sym}(x_{1},t_{1})>_{0} <T\mbox{ }\rho_{sym}(-x,t) \rho_{sym}(x',t')>_{0} }\nonumber\\& \mbox{ }\mbox{ }\substack{+ \mbox{ }<T\mbox{ }\rho_{sym}(-x',t') \rho_{sym}(x_{1},t_{1})>_{0} <T\mbox{ }\rho_{sym}(-x,t) \rho_{sym}(x_{2},t_{2})>_{0} <T\mbox{ }\rho_{sym}(x,t) \rho_{sym}(x',t')>_{0}}\nonumber\\& \mbox{ }\mbox{ }\substack{+ \mbox{ } <T\mbox{ }\rho_{sym}(-x',t') \rho_{sym}(x_{2},t_{2})>_{0} <T\mbox{ }\rho_{sym}(-x,t) \rho_{sym}(x_{1},t_{1})>_{0} <T\mbox{ }\rho_{sym}(x,t) \rho_{sym}(x',t')>_{0}}\bigg)
\end{align}

In momentum and Matsubara frequency space the second order term takes the form,
\begin{align}
<&\rho_{sym}(q',n)\rho_{sym}(q'',-n) >^{(2)}\nonumber\\&\mbox{ }=\mbox{ }\frac{(-i v_{0})^{2}}{32}(-i \beta)^{2}\mbox{ }\sum_{q,Q} \bigg(<\rho_{sym}(q,-n) \rho_{sym}(q',n) >_{0}\mbox{ }< \rho_{sym}(Q,n)\rho_{sym}(q'',-n) >_{0}\nonumber\\&\mbox{ }\mbox{ }\mbox{ }\mbox{ }\mbox{ }<\rho_{sym}(q,n)\rho_{sym}(Q,-n) >_{0} \mbox{ }+.......\bigg)
\end{align}
Using the expression for the non-interacting correlation in Fourier space
\begin{align*}
<\rho_{sym}(q,n)\rho_{sym}(q',-n)>_{0} = -\frac{1}{\pi \beta v_{F}}\frac{  i v_F q}{(w_{n}-i v_{F} q)} \delta_{q+q',0}
\end{align*}
we evaluate the second order term and obtain
\begin{align}
  <\rho_{sym}(q',n)\rho_{sym}(q'',-n) >^{(2)}\mbox{ }=\mbox{ }\frac{q^{'3} v_{0}^2}{4 \pi^3 \beta  (q' v_{F}-i w_{n}) (q' v_{F}+i w_{n})^2}\mbox{ }\delta_{q'+q'',0}
  \label{appendix2ordersym}
  \end{align}
Upon expanding Eq.\ref{rhosymmom1} in powers of $v_{0}$, the $O(v_{0}^{2})$ term is equal to Eq.\ref{appendix2ordersym}.

\section*{APPENDIX  B: Perturbative comparison of second order ($O(v_{0}^{2})$) term for $<\rho_{asy}\rho_{asy}>$}
\label{AppendixB}
\setcounter{equation}{0}
\renewcommand{\theequation}{B.\arabic{equation}}

The second order term in the expansion $<T \mbox{ } \rho_{asy}(x_{1},t_{1}) \rho_{asy}(x_{2},t_{2}) >$ is
\begin{align}
<&T \mbox{ } \rho_{asy}(x_{1},t_{1}) \rho_{asy}(x_{2},t_{2}) >^{(2)}\mbox{ }\nonumber\\&=\mbox{ }\substack{\frac{(-i v_{0})^{2}}{32} \int_{C}dt \int_{C} dt' \int dx \int dx' <T \mbox{ }\rho_{asy}(x,t)\rho_{asy}(-x,t) \rho_{asy}(x',t') \rho_{asy}(-x',t') \rho_{asy}(x_{1},t_{1}) \rho_{asy}(x_{2},t_{2})>_{0,c}}
\end{align}
Retaining the most singular parts alone we may write,
\begin{align}
&<T \mbox{ } \rho_{asy}(x_{1},t_{1}) \rho_{asy}(x_{2},t_{2})>^{(2)} \mbox{ }\nonumber\\&=\mbox{ } \substack{\frac{(-i v_{0})^{2}}{32} \int_{C} dt \int_{C} dt' \int dx \int dx' <T \mbox{ } \rho_{asy}(x,t)\rho_{asy}(-x,t) \rho_{asy}(x',t') \rho_{asy}(-x',t') \rho_{asy}(x_{1},t_{1}) \rho_{asy}(x_{2},t_{2}) >_{0,c}}\nonumber\\
& = \substack{\frac{(-i v_{0})^{2}}{32} \int_{C} dt \int_{C} dt' \int dx \int dx' \bigg( <T\mbox{ }\rho_{asy}(x,t) \rho_{asy}(x_{1},t_{1})>_{0} <T\mbox{ }\rho_{asy}(x',t') \rho_{asy}(x_{2},t_{2})>_{0} <T\mbox{ }\rho_{asy}(-x,t) \rho_{asy}(-x',t')>_{0}}\nonumber\\&\mbox{ }\mbox{ }\substack{+ \mbox{ }<T\mbox{ }\rho_{asy}(x,t) \rho_{asy}(x_{2},t_{2})>_{0} <T\mbox{ }\rho_{asy}(x',t') \rho_{asy}(x_{1},t_{1})>_{0} <T\mbox{ }\rho_{asy}(-x,t) \rho_{asy}(-x',t')>_{0}}\nonumber\\& \mbox{ }\mbox{ }\substack{+ \mbox{ }<T\mbox{ }\rho_{asy}(-x,t) \rho_{asy}(x_{1},t_{1})>_{0} <T\mbox{ }\rho_{asy}(x',t') \rho_{asy}(x_{2},t_{2})>_{0} <T\mbox{ }\rho_{asy}(x,t) \rho_{asy}(-x',t')>_{0} }\nonumber\\& \mbox{ }\mbox{ }\substack{+ \mbox{ }<T\mbox{ }\rho_{asy}(-x,t) \rho_{asy}(x_{2},t_{2})>_{0} <T\mbox{ }\rho_{asy}(x',t') \rho_{asy}(x_{1},t_{1})>_{0} <T\mbox{ }\rho_{asy}(x,t) \rho_{asy}(-x',t')>_{0}}\nonumber\\& \mbox{ }\mbox{ }\substack{+ \mbox{ }<T\mbox{ }\rho_{asy}(-x',t') \rho_{asy}(x_{1},t_{1})>_{0} <T\mbox{ }\rho_{asy}(x,t) \rho_{asy}(x_{2},t_{2})>_{0} <T\mbox{ }\rho_{asy}(-x,t) \rho_{asy}(x',t')>_{0}}\nonumber\\& \mbox{ }\mbox{ }\substack{+\mbox{ }<T\mbox{ }\rho_{asy}(-x',t') \rho_{asy}(x_{2},t_{2})>_{0} <T\mbox{ }\rho_{asy}(x,t) \rho_{asy}(x_{1},t_{1})>_{0} <T\mbox{ }\rho_{asy}(-x,t) \rho_{asy}(x',t')>_{0} }\nonumber\\& \mbox{ }\mbox{ }\substack{+ \mbox{ }<T\mbox{ }\rho_{asy}(-x',t') \rho_{asy}(x_{1},t_{1})>_{0} <T\mbox{ }\rho_{asy}(-x,t) \rho_{asy}(x_{2},t_{2})>_{0} <T\mbox{ }\rho_{asy}(x,t) \rho_{asy}(x',t')>_{0}}\nonumber\\& \mbox{ }\mbox{ }\substack{+ \mbox{ } <T\mbox{ }\rho_{asy}(-x',t') \rho_{asy}(x_{2},t_{2})>_{0} <T\mbox{ }\rho_{asy}(-x,t) \rho_{asy}(x_{1},t_{1})>_{0} <T\mbox{ }\rho_{asy}(x,t) \rho_{asy}(x',t')>_{0}}\bigg)
\end{align}
This is written in Fourier space as
\begin{align}
<&\rho_{asy}(q',n)\rho_{asy}(q'',-n) >^{(2)}\nonumber\\&\mbox{ }=\mbox{ }\frac{(-i v_{0})^{2}}{32}(-i \beta)^{2}\mbox{ }\sum_{q,Q} \bigg(<\rho_{asy}(q,-n) \rho_{asy}(q',n) >_{0}\mbox{ }< \rho_{asy}(Q,n)\rho_{asy}(q'',-n) >_{0}\nonumber\\&\mbox{ }\mbox{ }\mbox{ }\mbox{ }\mbox{ }<\rho_{asy}(q,n)\rho_{asy}(Q,-n) >_{0} \mbox{ }+.......\bigg)
\end{align}
Making use of
\begin{align}
< \rho_{asy}(q,n)\rho_{asy}(q',-n) >_{0} \mbox{ }=\mbox{ } -\frac{(i q v_{F}) \delta _{q+q',0}}{(\pi  \beta  v_{F}) (w_{n}-i q v_{F})}-\frac{(r_{1}-1) w_{n} sgn(w_{n})}{\pi  \beta  L (q v_{F}+i w_{n}) (q' v_{F}-i w_{n})}
\end{align}
we evaluate the second order term and obtain the following expression,
\begin{align}
<  &\rho_{asy}(q',n) \rho_{asy}(q'',-n) >^{(2)}\mbox{ }\nonumber\\&=\mbox{ }v_{0}^2 \bigg(-\frac{\substack{(r_{1}-1) w_{n} sgn(w_{n})} \bigg(\substack{q' \left(q'' v_{F} \left(\left(r_{1}^2-1\right) v_{F} v_{h}+(r_{1}+1)^2 v_{F}^2-(r_{1}-1)^2 v_{h}^2\right)+i (r_{1}-1) v_{h} w_{n} (2 (r_{1}+1) v_{F}-r_{1} v_{h}+v_{h})\right)\\+(r_{1}-1) v_{h} w_{n} ((r_{1}+1) w_{n}-i q'' (2 (r_{1}+1) v_{F}-r_{1} v_{h}+v_{h}))\bigg)}}{\substack{2 \pi ^3 \beta  L \left(q'^2 v_{h}^2+w_{n}^2\right) \left(q''^{2} v_{h}^2+w_{n}^2\right) (r_{1} v_{F}-r_{1} v_{h}+v_{F}+v_{h})^3}}\nonumber \\&\mbox{ }\mbox{ }\mbox{ }\mbox{ }\mbox{ }\mbox{ }+\substack{\frac{q'^3 \delta _{q',-q''}}{4 \pi^3 \beta  (q' v_{F}-i w_{n}) (q' v_{F}+i w_{n})^2}}\bigg)
\end{align}
 This is exactly equal to the second order term obtained upon perturbative expansion of the $<  \rho_{asy}(q',n) \rho_{asy}(q'',-n) >$ correlation function (present in the rhs of Eq.\ref{rhoasyrhoasyft}) in powers of $v_{0}$. 

\section*{APPENDIX  C: Validity of the Gaussian approximation}
\label{AppendixC}
\setcounter{equation}{0}
\renewcommand{\theequation}{C.\arabic{equation}}
In Eq.\ref{z0u} only the quadratic moment of the density is included in the non-interacting generating functional $Z_{0}[U']$, while the higher order moments are neglected. This Gaussian approximation is valid provided one is interested in only the most singular parts of the correlations, the precise meaning of this is clarified in this section. For a translationally invariant Luttinger liquid, the N-point correlation functions are nothing but the summation of the RPA diagrams \cite{dzyaloshinskii1974correlation,giamarchi2003quantum} that are written in powers of the mutual coupling between the fermions. But for a system with broken translational invariance such as the one considered in the present work (point-contact impurity) the Gaussian approximation should be interpreted as the resummation of the most singular RPA-level diagrams written in powers of the mutual coupling between the fermions but with the impurity treated exactly and this idea is part of already existing literature on this subject \cite{Danny_Babu_2020,das2019nonchiral,doi:10.1142/S0217751X18501749,Das_2018}. We carefully define the precise meaning of the \textit{most singular} term here.\\
The density fields are defined as the density fluctuations from the average and all higher order odd connected moments of the density fluctuation are always zero for a linear dispersion. When the Gaussian approximation is exact this means that all the higher order even connected density correlation functions also vanish. This is always true in the homogeneous case (without impurity), but in the presence of an arbitrary impurity the higher order even moments of the density are not necessarily zero but turn out to be less singular than the second moment. As long as one is interested in only the most singular part of the correlations one may choose to omit the less singular contributions. Let us consider the connected quartic moment of the density fields. When the Gaussian approximation is exact it means that the connected four-point density correlation function is
\begin{align}
< T& \mbox{  } \rho_{\nu_1}(x_1,t_1)\rho_{\nu_2}(x_2,t_2)\rho_{\nu_3}(x_3,t_3)\rho_{\nu_4}(x_4,t_4) >_{0,c}\nonumber\\ 
= &\mbox{ }< T \mbox{  } \rho_{\nu_1}(x_1,t_1)\rho_{\nu_2}(x_2,t_2)\rho_{\nu_3}(x_3,t_3)\rho_{\nu_4}(x_4,t_4) >_0 \nonumber\\
&- < T \mbox{  } \rho_{\nu_1}(x_1,t_1)\rho_{\nu_2}(x_2,t_2)>_{0}<\rho_{\nu_3}(x_3,t_3)\rho_{\nu_4}(x_4,t_4) >_0  \nonumber\\
&-< T \mbox{  } \rho_{\nu_1}(x_1,t_1)\rho_{\nu_3}(x_3,t_3)>_{0}<\rho_{\nu_2}(x_2,t_2)\rho_{\nu_4}(x_4,t_4) >_0  \nonumber\\
&-< T \mbox{  } \rho_{\nu_1}(x_1,t_1)\rho_{\nu_4}(x_4,t_4)>_{0}<\rho_{\nu_2}(x_2,t_2)\rho_{\nu_3}(x_3,t_3) >_0  \nonumber \\
= &\mbox{ }0
\end{align}
where $\nu_{1}, \nu_{2}, \nu_{3}\mbox{ }\text{and}\mbox{ }\nu_{4}$ denote $R$ or $L$ for right or left movers respectively. It follows that all higher order even moments also vanish when the Gaussian approximation is exact. We use Wick's theorem to express the general 4-density correlation function,
\small
\begin{align}
< T \mbox{  } &\rho_{\nu_1}(x_1,t_1)\rho_{\nu_2}(x_2,t_2)\rho_{\nu_3}(x_3,t_3)\rho_{\nu_4}(x_4,t_4) >_0  \nonumber \\\mbox{  }
=& <T \mbox{  } \psi_{\nu_1}^{\dagger}(x_1,t_1)\psi_{\nu_1}(x_1,t_1)\psi_{\nu_2}^{\dagger}(x_2,t_2)\psi_{\nu_2}(x_2,t_2)
\psi_{\nu_3}^{\dagger}(x_3,t_3)\psi_{\nu_3}(x_3,t_3) \psi_{\nu_4}^{\dagger}(x_4,t_4)\psi_{\nu_4}(x_4,t_4)>_{0}\nonumber \\
=& <T \mbox{  } \psi_{\nu_1}(x_1,t_1)\psi_{\nu_2}^{\dagger}(x_2,t_2)>_{0}<T\mbox{ }\psi_{\nu_2}(x_2,t_2)\psi_{\nu_1}^{\dagger}(x_1,t_1)>_{0}
<T\mbox{ }\psi_{\nu_3}(x_3,t_3) \psi_{\nu_4}^{\dagger}(x_4,t_4)>_{0}\nonumber\\ &<T\mbox{ }\psi_{\nu_4}(x_4,t_4)\psi_{\nu_3}^{\dagger}(x_3,t_3)>_{0}\nonumber \\
&+ <T \mbox{  } \psi_{\nu_1}(x_1,t_1)\psi_{\nu_3}^{\dagger}(x_3,t_3)>_{0}<T\mbox{ }\psi_{\nu_2}(x_2,t_2)\psi_{\nu_4}^{\dagger}(x_4,t_4)>_{0}
<T\mbox{ }\psi_{\nu_3}(x_3,t_3) \psi_{\nu_1}^{\dagger}(x_1,t_1)>_{0}\nonumber \\ &<T\mbox{ }\psi_{\nu_4}(x_4,t_4)\psi_{\nu_2}^{\dagger}(x_2,t_2)>_{0}\nonumber\\
&+<T \mbox{  } \psi_{\nu_1}(x_1,t_1)\psi_{\nu_4}^{\dagger}(x_4,t_4)>_{0}<T\mbox{ }\psi_{\nu_2}(x_2,t_2)\psi_{\nu_3}^{\dagger}(x_3,t_3)>_{0}
<T\mbox{ }\psi_{\nu_3}(x_3,t_3) \psi_{\nu_2}^{\dagger}(x_2,t_2)>_{0}\nonumber \\ &<T\mbox{ }\psi_{\nu_4}(x_4,t_4)\psi_{\nu_1}^{\dagger}(x_1,t_1)>_{0}\nonumber \\
&- <T \mbox{  } \psi_{\nu_1}(x_1,t_1)\psi_{\nu_2}^{\dagger}(x_2,t_2)>_{0}<T\mbox{ }\psi_{\nu_2}(x_2,t_2)\psi_{\nu_3}^{\dagger}(x_3,t_3)>_{0}
<T\mbox{ }\psi_{\nu_3}(x_3,t_3) \psi_{\nu_4}^{\dagger}(x_4,t_4)>_{0}\nonumber \\ &<T\mbox{ }\psi_{\nu_4}(x_4,t_4)\psi_{\nu_1}^{\dagger}(x_1,t_1)>_{0}\nonumber \\
&- <T \mbox{  } \psi_{\nu_1}(x_1,t_1)\psi_{\nu_2}^{\dagger}(x_2,t_2)>_{0}<T\mbox{ }\psi_{\nu_2}(x_2,t_2)\psi_{\nu_4}^{\dagger}(x_4,t_4)>_{0}
<T\mbox{ }\psi_{\nu_3}(x_3,t_3) \psi_{\nu_1}^{\dagger}(x_1,t_1)>_{0}\nonumber \\ &<T\mbox{ }\psi_{\nu_4}(x_4,t_4)\psi_{\nu_3}^{\dagger}(x_3,t_3)>_{0}\nonumber \\
&- <T \mbox{  } \psi_{\nu_1}(x_1,t_1)\psi_{\nu_3}^{\dagger}(x_3,t_3)>_{0}<T\mbox{ }\psi_{\nu_2}(x_2,t_2)\psi_{\nu_1}^{\dagger}(x_1,t_1)>_{0}
<T\mbox{ }\psi_{\nu_3}(x_3,t_3) \psi_{\nu_4}^{\dagger}(x_4,t_4)>_{0}\nonumber \\ &<T\mbox{ }\psi_{\nu_4}(x_4,t_4)\psi_{\nu_2}^{\dagger}(x_2,t_2)>_{0}\nonumber \\
&- <T \mbox{  } \psi_{\nu_1}(x_1,t_1)\psi_{\nu_3}^{\dagger}(x_3,t_3)>_{0}<T\mbox{ }\psi_{\nu_2}(x_2,t_2)\psi_{\nu_4}^{\dagger}(x_4,t_4)>_{0}
<T\mbox{ }\psi_{\nu_3}(x_3,t_3) \psi_{\nu_2}^{\dagger}(x_2,t_2)>_{0}\nonumber \\ &<T\mbox{ }\psi_{\nu_4}(x_4,t_4)\psi_{\nu_1}^{\dagger}(x_1,t_1)>_{0}\nonumber \\
&- <T \mbox{  } \psi_{\nu_1}(x_1,t_1)\psi_{\nu_4}^{\dagger}(x_4,t_4)>_{0}<T\mbox{ }\psi_{\nu_2}(x_2,t_2)\psi_{\nu_1}^{\dagger}(x_1,t_1)>_{0}
<T\mbox{ }\psi_{\nu_3}(x_3,t_3) \psi_{\nu_2}^{\dagger}(x_2,t_2)>_{0}\nonumber \\ &<T\mbox{ }\psi_{\nu_4}(x_4,t_4)\psi_{\nu_3}^{\dagger}(x_3,t_3)>_{0}\nonumber \\
&- <T \mbox{  } \psi_{\nu_1}(x_1,t_1)\psi_{\nu_4}^{\dagger}(x_4,t_4)>_{0}<T\mbox{ }\psi_{\nu_2}(x_2,t_2)\psi_{\nu_3}^{\dagger}(x_3,t_3)>_{0}
<T\mbox{ }\psi_{\nu_3}(x_3,t_3) \psi_{\nu_1}^{\dagger}(x_1,t_1)>_{0}\nonumber \\ &<T\mbox{ }\psi_{\nu_4}(x_4,t_4)\psi_{\nu_2}^{\dagger}(x_2,t_2)>_{0}
\label{generalfourmoment}
\end{align}
\normalsize
The non-interacting two-point Green's function at equilibrium for the present model is given by Eq.13 in \cite{Babu_2022} with zero bias voltage. We have (at zero temperature for simplicity)
\begin{align}
\langle T\mbox{  }\psi_{\nu}(x,t)\psi_{\nu'}^{\dagger}(x',t')\rangle_0 \mbox{ }\sim\mbox{ } O\left(\frac{1}{(\nu x-\nu' x')-v_F(t-t')}\right)
\end{align}
where $O$ represents the order. Thus we may write
\begin{align}
< T \mbox{  } &\rho_{\nu_1}(x_1,t_1)\rho_{\nu_2}(x_2,t_2)\rho_{\nu_3}(x_3,t_3)\rho_{\nu_4}(x_4,t_4) >_0  \nonumber\\\mbox{  }
=& <T \mbox{  } \psi_{\nu_1}^{\dagger}(x_1,t_1)\psi_{\nu_1}(x_1,t_1)\psi_{\nu_2}^{\dagger}(x_2,t_2)\psi_{\nu_2}(x_2,t_2)
\psi_{\nu_3}^{\dagger}(x_3,t_3)\psi_{\nu_3}(x_3,t_3) \psi_{\nu_4}^{\dagger}(x_4,t_4)\psi_{\nu_4}(x_4,t_4)>_{0} \nonumber \\
=&\mbox{ }O\left(\frac{1}{(\nu_1 x_1-\nu_2x_2-v_F(t_1-t_2))^2(\nu_3 x_3-\nu_4x_4-v_F(t_3-t_4))^2}  \right) \nonumber \\
&+O\left(\frac{1}{(\nu_1 x_1-\nu_3x_3-v_F(t_1-t_3))^2(\nu_2 x_2-\nu_4x_4-v_F(t_2-t_4))^2}  \right) \nonumber \\
&+O\left(\frac{1}{(\nu_1 x_1-\nu_4x_4-v_F(t_1-t_4))^2(\nu_2 x_2-\nu_3x_3-v_F(t_2-t_3))^2}  \right) \nonumber \\
&-O\left( \frac{1}{\substack{(\nu_1 x_1-\nu_2x_2-v_F(t_1-t_2))(\nu_2 x_2-\nu_3x_3-v_F(t_2-t_3))(\nu_3 x_3-\nu_4x_4-v_F(t_3-t_4))(\nu_4 x_4-\nu_1x_1-v_F(t_4-t_1))}} \right) \nonumber \\
&-O\left( \frac{1}{\substack{(\nu_1 x_1-\nu_2x_2-v_F(t_1-t_2))(\nu_2 x_2-\nu_4x_4-v_F(t_2-t_4))(\nu_3 x_3-\nu_1x_1-v_F(t_3-t_1))(\nu_4 x_4-\nu_3x_3-v_F(t_4-t_3))}} \right) \nonumber \\
&-O\left( \frac{1}{\substack{(\nu_1 x_1-\nu_3x_3-v_F(t_1-t_3))(\nu_2 x_2-\nu_1x_1-v_F(t_2-t_1))(\nu_3 x_3-\nu_4x_4-v_F(t_3-t_4))(\nu_4 x_4-\nu_2x_2-v_F(t_4-t_2))}} \right) \nonumber \\
&-O\left( \frac{1}{\substack{(\nu_1 x_1-\nu_3x_3-v_F(t_1-t_3))(\nu_2 x_2-\nu_4x_4-v_F(t_2-t_4))(\nu_3 x_3-\nu_2x_2-v_F(t_3-t_2))(\nu_4 x_4-\nu_1x_1-v_F(t_4-t_1))}} \right) \nonumber \\
&-O\left( \frac{1}{\substack{(\nu_1 x_1-\nu_4x_4-v_F(t_1-t_4))(\nu_2 x_2-\nu_1x_1-v_F(t_2-t_1))(\nu_3 x_3-\nu_2x_2-v_F(t_3-t_2))(\nu_4 x_4-\nu_3x_3-v_F(t_4-t_3))}} \right) \nonumber \\
&-O\left( \frac{1}{\substack{(\nu_1 x_1-\nu_4x_4-v_F(t_1-t_4))(\nu_2 x_2-\nu_3x_3-v_F(t_2-t_3))(\nu_3 x_3-\nu_1x_1-v_F(t_3-t_1))(\nu_4 x_4-\nu_2x_2-v_F(t_4-t_2))}} \right)
\label{ffunction}
\end{align}
The first three terms in the above expression are more singular than the rest of the terms. The notion of the \textit{most singular part} of an expression can be made sense of in the following manner. The correlation functions can be thought of as functions of the time difference $\tau = t_{1}-t_{2}$ (which they are). When we have terms of the form (for e.g.)
\begin{align}
\frac{C}{(\tau - a)^{2}} + \frac{D}{(\tau - c1)(\tau-c2)}
\end{align}
The first term is considered more singular than the second (if $c1 \neq c2$) since the former is a second order pole whereas the latter when partial fraction expanded is a sum of two first order poles. Calculating transport properties of interest like the tunneling current or conductance through the point-contact involves evaluating the Green's functions in the equal space-equal time limits. It is the \textit{most singular parts} of the correlations that become important in these limits. We can write
\begin{align}
< T \mbox{  }& \rho_{\nu_1}(x_1,t_1)\rho_{\nu_2}(x_2,t_2)><\rho_{\nu_3}(x_3,t_3)\rho_{\nu_4}(x_4,t_4) >_0  
+< T \mbox{  } \rho_{\nu_1}(x_1,t_1)\rho_{\nu_3}(x_3,t_3)><\rho_{\nu_2}(x_2,t_2)\rho_{\nu_4}(x_4,t_4) >_0 \nonumber \\
+&< T \mbox{  } \rho_{\nu_1}(x_1,t_1)\rho_{\nu_4}(x_4,t_4)><\rho_{\nu_2}(x_2,t_2)\rho_{\nu_3}(x_3,t_3) >_0  \nonumber\\
=& <T \mbox{  } \psi_{\nu_1}(x_1,t_1)\psi_{\nu_2}^{\dagger}(x_2,t_2)>_{0}<T\mbox{ }\psi_{\nu_2}(x_2,t_2)\psi_{\nu_1}^{\dagger}(x_1,t_1)>_{0}\nonumber \\
&<T\mbox{ }\psi_{\nu_3}(x_3,t_3) \psi_{\nu_4}^{\dagger}(x_4,t_4)>_{0}<T\mbox{ }\psi_{\nu_4}(x_4,t_4)\psi_{\nu_3}^{\dagger}(x_3,t_3)>_{0}\nonumber\\
&+ <T \mbox{  } \psi_{\nu_1}(x_1,t_1)\psi_{\nu_3}^{\dagger}(x_3,t_3)>_{0}<T\mbox{ }\psi_{\nu_2}(x_2,t_2)\psi_{\nu_4}^{\dagger}(x_4,t_4)>_{0}\nonumber \\
&<T\mbox{ }\psi_{\nu_3}(x_3,t_3) \psi_{\nu_1}^{\dagger}(x_1,t_1)>_{0}<T\mbox{ }\psi_{\nu_4}(x_4,t_4)\psi_{\nu_2}^{\dagger}(x_2,t_2)>_{0}\nonumber\\
&+ <T \mbox{  } \psi_{\nu_1}(x_1,t_1)\psi_{\nu_4}^{\dagger}(x_4,t_4)>_{0}<T\mbox{ }\psi_{\nu_2}(x_2,t_2)\psi_{\nu_3}^{\dagger}(x_3,t_3)>_{0}\nonumber \\
&<T\mbox{ }\psi_{\nu_3}(x_3,t_3) \psi_{\nu_2}^{\dagger}(x_2,t_2)>_{0}<T\mbox{ }\psi_{\nu_4}(x_4,t_4)\psi_{\nu_1}^{\dagger}(x_1,t_1)>_{0}
\end{align}
Hence we may write Eq.\ref{ffunction} as 
\begin{align}
< T \mbox{  }& \rho_{\nu_1}(x_1,t_1)\rho_{\nu_2}(x_2,t_2)\rho_{\nu_3}(x_3,t_3)\rho_{\nu_4}(x_4,t_4) >_0  \mbox{  }
\nonumber \\= &< T \mbox{  } \rho_{\nu_1}(x_1,t_1)\rho_{\nu_2}(x_2,t_2)><\rho_{\nu_3}(x_3,t_3)\rho_{\nu_4}(x_4,t_4) >_0  \nonumber\\
&+< T \mbox{  } \rho_{\nu_1}(x_1,t_1)\rho_{\nu_3}(x_3,t_3)><\rho_{\nu_2}(x_2,t_2)\rho_{\nu_4}(x_4,t_4) >_0  \nonumber\\
&+< T \mbox{  } \rho_{\nu_1}(x_1,t_1)\rho_{\nu_4}(x_4,t_4)><\rho_{\nu_2}(x_2,t_2)\rho_{\nu_3}(x_3,t_3) >_0 \nonumber \\
&+ \mbox{ }\text{less singular terms}
\end{align}
This means that the connected 4-density correlation function is
\begin{align}
< T \mbox{  }& \rho_{\nu_1}(x_1,t_1)\rho_{\nu_2}(x_2,t_2)\rho_{\nu_3}(x_3,t_3)\rho_{\nu_4}(x_4,t_4) >_{0,c} = \text{less singular terms}
\end{align}
Note that the quartic moments of the $\rho_{sym}$ and $\rho_{asy}$ fields will involve correlations precisely of this form (these fields are defined in terms of $\rho_{R}$ and $\rho_{L}$ in Eq.\ref{rsymnonint}). The expression for the two-point correlations were obtained in Eq.13 of reference \cite{Babu_2022} which in equilibrium is written as (at zero temperature)
 \begin{align}
 <\psi^{\dagger}_{\nu^{'}}(x^{'},t^{'})\psi_{\nu}(x,t)> \mbox{ }=\mbox{ } -\frac{i}{2\pi} \frac{1}{ (\nu x-\nu^{'}x^{'}-v_F(t-t^{'}) )} \kappa_{\nu,\nu^{'}}
 \label{twopointgreen}
 \end{align}
 where $\nu$,$\nu^{'} = \pm 1$ with $R=1$ and $L=-1$ and
 \begin{widetext}
 \small
 \begin{align}
 \kappa_{1,1} \mbox{ }=\mbox{ } \bigg( \left[1
-      \theta(x^{'})    \mbox{          }   \frac{  2 \Gamma^2
}{\Gamma ^2 +4 v_F^2}\right]&\mbox{          }    \left[1
-      \theta(x )    \mbox{          }   \frac{  2 \Gamma^2
}{\Gamma ^2 +4 v_F^2}\right]
  \nonumber \\ &+ \left( \frac{\Gamma }{v_F} \mbox{          } \frac{(2 v_F)^2
}{\Gamma ^2 +4 v_F^2}\right)^2 \mbox{          }
  \theta(x )     \theta(x^{'} )   \mbox{          } \bigg)
   \label{eqkappa1}
 \end{align}
 \begin{align}
 \kappa_{-1,-1} \mbox{ }=\mbox{ } \bigg( \left[ 1
-      \theta(-x^{'})    \mbox{          }   \frac{  2 \Gamma^2
}{\Gamma ^2 +4 v_F^2}
\right] &  \left[ 1
-      \theta(-x )    \mbox{          }   \frac{  2 \Gamma^2
}{\Gamma ^2 +4 v_F^2}
\right]   \nonumber \\ &+   \left( \frac{\Gamma }{v_F} \mbox{          } \frac{(2 v_F)^2
}{\Gamma ^2 +4 v_F^2}\right)^2 \mbox{          }  \theta(-x )    \theta( -x^{'} )    \mbox{          }
 \bigg)
 \label{eqkappa2}
 \end{align}
 \begin{align}
 \kappa_{1,-1} \mbox{ }= \mbox{ }\bigg( - \mbox{          }  &\left[ 1
-      \theta(-x^{'})    \mbox{          }   \frac{  2 \Gamma^2
}{\Gamma ^2 +4 v_F^2}
\right]
    \theta(x )   \mbox{          } \nonumber \\ &+
\mbox{          }
\left[1
-      \theta(x )    \mbox{          }   \frac{  2 \Gamma^2
}{\Gamma ^2 +4 v_F^2}\right] \theta( -x^{'} )   \bigg) i  \frac{\Gamma }{v_F}\frac{(2 v_F)^2
}{\Gamma ^2 +4 v_F^2}
\label{eqkappa3}
 \end{align}
 \begin{align}
 \kappa_{-1,1} \mbox{ }=\mbox{ }\bigg( -  &\left[1
-      \theta(x^{'})    \mbox{          }   \frac{  2 \Gamma^2
}{\Gamma ^2 +4 v_F^2}\right]  \theta( -x )
 \nonumber \\ &+       \mbox{          }     \left[ 1
-      \theta(-x )    \mbox{          }   \frac{  2 \Gamma^2
}{\Gamma ^2 +4 v_F^2}
\right]  \theta(x^{'} )     \bigg) \mbox{          } i \frac{\Gamma }{v_F}    \frac{ (  2  v_F )^2   }{\Gamma ^2 +4 v_F^2}
\label{eqkappa4}
 \end{align}
 \normalsize
 \end{widetext}
It is easy to check (using Eqs.\ref{twopointgreen} to \ref{eqkappa4}) that the odd order density correlations are always trivially zero. For example the three-point density correlations are,
\begin{align}
< T&\mbox{ }\rho_{\nu_1}(x_1,t_1) \rho_{\nu_2}(x_2,t_2) \rho_{\nu_3}(x_3,t_3)>_{0,c}\nonumber \\
 = & \mbox{ }<T\mbox{   }\psi^{\dagger}_{\nu_1}(x_1,t_1)\psi_{\nu_1}(x_1,t_1) \psi^{\dagger}_{\nu_2}(x_2,t_2)\psi_{\nu_2}(x_2,t_2) \psi^{\dagger}_{\nu_3}(x_3,t_3)\psi_{\nu_3}(x_3,t_3)>_{0,c}\nonumber \\
 = & \mbox{ }-<T\mbox{   }\psi_{\nu_2}(x_2,t_2)\psi^{\dagger}_{\nu_1}(x_1,t_1)>_0 <T \mbox{  }\psi_{\nu_1}(x_1,t_1)\psi^{\dagger}_{\nu_3}(x_3,t_3)>_0<T\mbox{    } \psi_{\nu_3}(x_3,t_3)\psi^{\dagger}_{\nu_2}(x_2,t_2) >_0\nonumber \\
& \mbox{ }- <T\mbox{   }\psi_{\nu_3}(x_3,t_3)\psi^{\dagger}_{\nu_1}(x_1,t_1)>_0<T\mbox{   } \psi_{\nu_1}(x_1,t_1) \psi^{\dagger}_{\nu_2}(x_2,t_2)>_0\mbox{   }<T\mbox{  }\psi_{\nu_2}(x_2,t_2) \psi^{\dagger}_{\nu_3}(x_3,t_3)>_0 \nonumber \\
= & \mbox{ }0 
\end{align}
All higher order odd moments of the density are zero even in the homogeneous (no impurity) case.
Using Wick's theorem Eq.\ref{generalfourmoment} it is easy to see that the fourth order connected moments in the homogeneous case ($\Gamma = 0$) are zero. As an example let us evaluate the following correlation function with all points on the same side ($x_{1}>0,\mbox{ }x_{2}>0,\mbox{ }x_{3}>0,\mbox{ }x_{4}>0$),
\begin{align}
< &T \mbox{  } \rho_{R}(x_1,t_1)\rho_{R}(x_2,t_2)\rho_{R}(x_3,t_3)\rho_{R}(x_4,t_4) >_{0,c}\nonumber\\
\mbox{ }=&\mbox{ }-\frac{1}{2 \pi^{4}}\bigg(\left( 1
-          \mbox{          }   \frac{  2 \Gamma^2
}{\Gamma ^2 +4 v_F^2} \right)^{2} + \left( \frac{\Gamma }{v_F} \mbox{          } \frac{(2 v_F)^2
}{\Gamma ^2 +4 v_F^2}\right)^2\bigg)^{4}\nonumber \\
& \bigg( \frac{1}{( x_1-x_2-v_F(t_1-t_2))( x_2-x_3-v_F(t_2-t_3))( x_3-x_4-v_F(t_3-t_4))( x_4-x_1-v_F(t_4-t_1))}  \nonumber \\
&+ \frac{1}{( x_1-x_2-v_F(t_1-t_2))( x_2-x_4-v_F(t_2-t_4))( x_3-x_1-v_F(t_3-t_1))( x_4-x_3-v_F(t_4-t_3))}  \nonumber \\
&+ \frac{1}{( x_1-x_3-v_F(t_1-t_3))( x_2-x_1-v_F(t_2-t_1))( x_3-x_4-v_F(t_3-t_4))( x_4-x_2-v_F(t_4-t_2))} \nonumber \\
&+ \frac{1}{( x_1-x_3-v_F(t_1-t_3))( x_2-x_4-v_F(t_2-t_4))( x_3-x_2-v_F(t_3-t_2))( x_4-x_1-v_F(t_4-t_1))}  \nonumber \\
&+ \frac{1}{( x_1-x_4-v_F(t_1-t_4))( x_2-x_1-v_F(t_2-t_1))( x_3-x_2-v_F(t_3-t_2))( x_4-x_3-v_F(t_4-t_3))} \nonumber \\
&+ \frac{1}{( x_1-x_4-v_F(t_1-t_4))( x_2-x_3-v_F(t_2-t_3))( x_3-x_1-v_F(t_3-t_1))( x_4-x_2-v_F(t_4-t_2))} \bigg)\nonumber \\
\mbox{ }=&\mbox{ }0
\end{align}
This simplifies to zero. As another example let us evaluate the following correlation function for the case ($x_{1}>0,\mbox{ }x_{2}>0, \mbox{ }x_{3}<0,\mbox{ }x_{4}<0$),
\begin{align}
< &T \mbox{  } \rho_{R}(x_1,t_1)\rho_{L}(x_2,t_2)\rho_{R}(x_3,t_3)\rho_{L}(x_4,t_4) >_{0,c}\nonumber\\
\mbox{ }=&\mbox{ }\frac{2(v_{F} \Gamma^{3} - 4 v_{F}^{3} \Gamma)^{2}}{\pi^{4}(4 v_{F}^{2} + \Gamma^{2})^{4}}\times\nonumber \\
& \bigg(\frac{1}{(x_{1}+x_{2} - v_{F}(t_{1}-t_{2}))(x_{1} - x_{3} - v_{F}(t_{1} - t_{3}))(-x_{2} + x_{4} - v_{F}(t_{2}-t_{4}))(x_{3} + x_{4} - v_{F}(t_{3}-t_{4}))}\bigg)
\end{align} 
It is clear that in the homogeneous limit ($\Gamma = 0$) this four-density function vanishes and in the inhomogeneous case this is non-zero but contains only first order poles and hence is less singular than the quadratic moment. Similarly it can be shown that all the four-density correlations are zero in the homogeneous case,
\begin{align}
< &T \mbox{  } \rho_{\nu_1}(x_1,t_1)\rho_{\nu_2}(x_2,t_2)\rho_{\nu_3}(x_3,t_3)\rho_{\nu_4}(x_4,t_4) >_{0,c}\nonumber \\
 =& < T \mbox{  } \rho_{\nu_1}(x_1,t_1)\rho_{\nu_2}(x_2,t_2)\rho_{\nu_3}(x_3,t_3)\rho_{\nu_4}(x_4,t_4) >_0  \mbox{  }
\nonumber \\&- < T \mbox{  } \rho_{\nu_1}(x_1,t_1)\rho_{\nu_2}(x_2,t_2)><\rho_{\nu_3}(x_3,t_3)\rho_{\nu_4}(x_4,t_4) >_0  \nonumber\\
&-< T \mbox{  } \rho_{\nu_1}(x_1,t_1)\rho_{\nu_3}(x_3,t_3)><\rho_{\nu_2}(x_2,t_2)\rho_{\nu_4}(x_4,t_4) >_0  \nonumber\\
&-< T \mbox{  } \rho_{\nu_1}(x_1,t_1)\rho_{\nu_4}(x_4,t_4)><\rho_{\nu_2}(x_2,t_2)\rho_{\nu_3}(x_3,t_3) >_0 \nonumber \\
&=\mbox{ }0
\end{align}
Consequently the higher order density correlations also vanish in this limit. Hence the Gaussian approximation is exact in the homogeneous case. In the inhomogeneous case i.e. in presence of an impurity all the higher order (beyond quadratic) moments are not necessarily zero but less singular than the quadratic moment and the less singular terms are omitted by enforcing the Gaussian approximation which is not exact in this case. It is only by making such an approximation (retaining the most singular parts) that it is possible to write down a compact analytical expression (Eqs.\ref{rsymrsymint} and \ref{rhoasyrhoasyst}) for the density-density correlation functions in presence of an arbitrary impurity and mutual interactions between the fermions. In fact, the density-density correlation functions (most singular part) that we have meticulously derived in this work have been used in conjunction with an unconventional bosonization procedure \cite{babu2023unconventional} to recover the correct tunneling density of states (TDOS) \cite{babu2023tunneling} for tunneling into a Luttinger liquid at the point-contact. In fact this idea of the most singular part is very familiar in many-body theory albeit in wave-vector/frequency space. To put this in perspective, let us take the familiar example of the Green's function for a system of mutually interacting particles in equilibrium assuming translational invariance in both space and time 
\begin{align*}
G({\bf{k}},z_{n})\mbox{ }=\mbox{ }\frac{1}{i z_{n} - \epsilon_{k} + \mu - \Sigma({\bf{k}},i z_{n})}
\end{align*}
where $\epsilon_{k} = \frac{\hbar^{2} k^{2}}{2m}$ and $\Sigma$ is the self-energy.
Here the Green's function is expressed in the momentum-Matsubara frequency space where $z_{n} = \frac{(2 n + 1)\pi}{\beta}$ for fermions and $z_{n} = \frac{2 n \pi}{\beta}$ for bosons with $n$ an integer. The zeros of the denominator of $G({\bf{k}},-i \omega)$ are the solutions of $\omega - \epsilon_{k} + \mu - \Sigma({\bf{k}},\omega) = 0$. We may express these solutions as $\omega\mbox{ }=\mbox{ }\omega_{1}({\bf{k}}) + i \omega_{2}({\bf{k}})$ where $\omega_{1,2}$ are real. This is essentially a pole and in the vicinity of the pole we may write the Green's function as (refer to Eq. 10.73 in \cite{setlur2013dynamics})
\begin{align*}
G({\bf{k}},-i \omega) \approx \frac{Z({\bf{k}})}{(\omega - \omega_{1}({\bf{k}}) - i \omega_{2}({\bf{k}}))} + G_{reg}({\bf{k}},-i \omega)
\end{align*}
where $G_{reg}$ is non-singular at the pole and $Z({\bf{k}})$ is the quasiparticle residue and it measures the jump in the momentum distribution across the Fermi surface for fermions and the condensate fraction for bosons. The first term in the above expression is the most singular part and $G_{reg}$ is non-singular and in the vicinity of the pole the more singular term is dominant. The most singular part of the correlations that we discuss in our work is analogous to this but in the space-time domain.

\section*{APPENDIX  D: Interpretation of the RPA}
\label{AppendixD}
\setcounter{equation}{0}
\renewcommand{\theequation}{D.\arabic{equation}}
In this work we are concerned with only the low energy properties of the system, that is we are interested only in the excitations close to the Fermi surface. To deal with this case we consider a linearized dispersion close to the Fermi points. We thus have two species of spinless fermions: the right and left movers. The kinetic energy part of the Hamiltonian is 
\begin{align}
H_{kin} = \sum_{p}(v_F p)c^{\dagger}_{p,R}c_{p,R} &+ \sum_{p}(-v_Fp)c^{\dagger}_{p,L}c_{p,L}
\end{align}
In the system under consideration, the opposite chiral edge states are coupled through an interaction mediated by the fractional quantum Hall bulk. The interaction Hamiltonian is
\begin{align}
H_{int} = \int dx\mbox{ } v_{0}\mbox{ } \rho_{R}(x,t) \rho_{L}(x,t)
\end{align}
where $v_{0}$ is the interaction parameter. The density fluctuation operator in momentum space is expressed as
\begin{align}
\rho^{\dagger}(q) = \sum_{k} c^{\dagger}_{k+q}c_{k}
\end{align}
with both $k$ and $k+q$ close to the Fermi points at $-k_{F}$ or $+k_{F}$ which means that interaction processes close to the Fermi surface is only of interest. The interaction term in our Hamiltonian couples fermions on one edge with fermions on the opposite edge, in other words it couples fermions on one side of the Fermi surface with fermions on the other side. However, each species of fermion remains on the same edge after the interaction i.e. there is no backscattering of fermions due to the interaction and the chirality of the fermions remains unchanged. This is equivalent to the $g2$ process in what is known as the g-ology terminology in the literature \cite{giamarchi2003quantum}. Since our model is spinless this is the only low energy interaction process of importance. It was shown by Dzyaloshinskii and Larkin \cite{dzyaloshinskii1974correlation} that in such a situation in a diagrammatic expansion of the interaction only diagrams with fermion bubbles with at most two interaction lines contribute. Such diagrams which are also called 2-loops are the two-point density-density correlation functions \cite{PhysRevB.58.15449}. Quite remarkably all the other bubble diagrams with more than two interaction lines cancel as a consequence of the linear dispersion. This means that contributions from N-loop diagrams (N-point density-density correlation functions) with $N > 2$ is zero. 
\begin{figure}[h!]
\centering
 \includegraphics[scale=0.2]{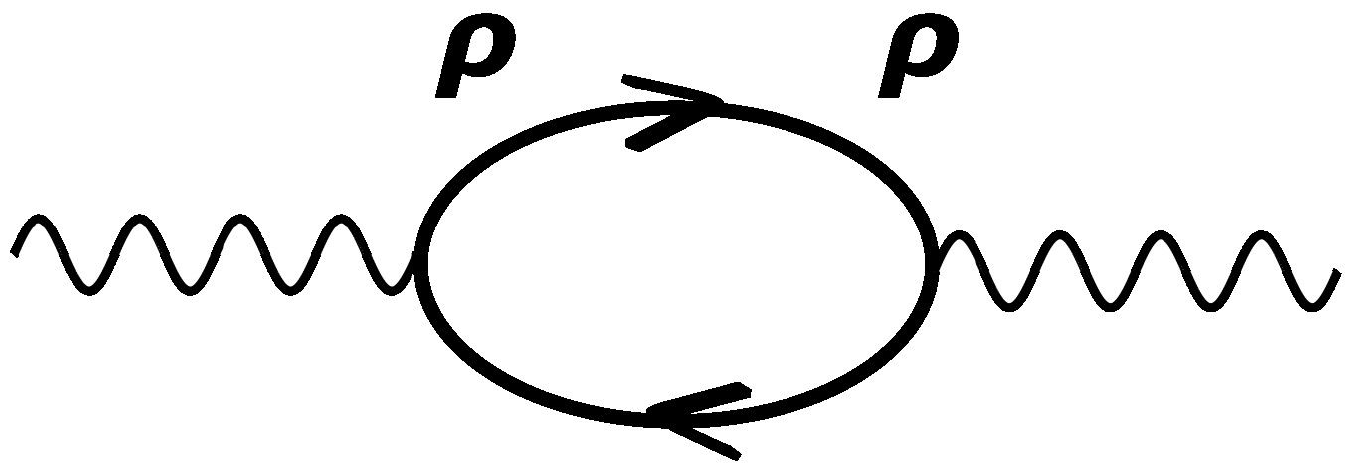}
 \caption{\small Only fermion bubbles with two interaction lines are present for a model with linear dispersion. Such diagrams correspond to the dynamical density-density correlation functions.}
 \label{bubbleplot}
 \end{figure}
Any given diagram will contain only bubbles or series of bubbles and a summation of such terms is what is known as the random phase approximation (RPA) \cite{PhysRev.82.625}. In the limit that the RPA is valid the Fermi momentum and mass of the fermion is allowed to diverge in such a way that their ratio is finite ($k_{F},\mbox{ }m \rightarrow \infty$ but $k_{F}/m = v_{F} < \infty$). Dzyaloshinskii's work \cite{dzyaloshinskii1974correlation} shows that the RPA is exact for the present problem with a linear dispersion in the homogeneous case. However, in presence of an impurity such as the point-contact in the present problem there is backscattering (change of chirality) of the fermions and the number of right movers and left movers is not separately conserved. In this case the N-loop diagrams with $N > 2$ do not cancel and the higher order density correlations are non-zero. In principle it is not possible to write down a compact analytical expression for the interacting density-density correlations. But as we have shown in \hyperref[AppendixC]{Appendix C} the higher even order density correlations are less singular than the two-point density correlation function (the odd order density correlations vanish for a linearized dispersion \cite{dzyaloshinskii1974correlation}). It is possible to obtain a compact analytical expression for the interacting density correlations if we choose to retain only the most singular part and omit the less singular contribution. When we make the Gaussian approximation of $Z_{0}$ in Eq.\ref{z0u}, this choice corresponds to the RPA for the homogeneous case, but for the inhomogeneous case this corresponds to the most singular truncation of the RPA generating functional. Another way of saying this is that for the inhomogeneous system a most singular truncation and resummation of the RPA diagrams is performed. The caveat is that our answer is not the full story, it is only the most singular part of the density-density correlations. This is a reasonable approximation since it is the singular behaviour of the correlations that become important when evaluating transport properties like the current and conductance. 
\mbox{ }\\

\section*{References}
\bibliographystyle{iopart-num}
\bibliography{refddcf}
\end{document}